\documentclass[11pt]{article}

\usepackage[margin=1in]{geometry}
\usepackage{xcolor}
\usepackage{amsmath,amssymb,amsthm,mathtools}
\usepackage{bm}
\usepackage{hyperref}
\usepackage{enumitem}
\usepackage{algorithm}
\usepackage{algpseudocode}
\usepackage{graphicx}
\usepackage{tikz}
\usetikzlibrary{arrows.meta,positioning}
\usepackage[round,authoryear]{natbib}

\hypersetup{colorlinks=true,linkcolor=blue,citecolor=blue,urlcolor=blue}

\theoremstyle{plain}
\newtheorem{theorem}{Theorem}
\newtheorem{proposition}{Proposition}
\newtheorem{lemma}{Lemma}

\theoremstyle{definition}
\newtheorem{assumption}{Assumption}

\newtheorem{example}{Example}
\newtheorem{remark}{Remark}

\newcommand{\R}{\mathbb{R}}
\newcommand{\N}{\mathbb{N}}
\newcommand{\E}{\mathbb{E}}
\newcommand{\Prob}{\mathbb{P}}
\newcommand{\Psp}{\mathcal{P}}
\newcommand{\PspOne}{\mathcal{P}_1}
\newcommand{\Lip}{\mathrm{Lip}}
\newcommand{\TV}{\mathrm{TV}}

\newcommand{\dist}{d}
\newcommand{\Wone}{W_1}
\newcommand{\Xs}{\mathcal{X}}
\newcommand{\argmin}{\operatorname*{arg\,min}}
\newcommand{\ip}[2]{\langle #1, #2\rangle}
\newcommand{\thn}{\tilde{\theta}_{n_0}}
\newcommand{\bw}{\mathbf{w}}
\newcommand{\Pxo}{\widehat P_{\xi_0}}
\newcommand{\Pxi}{\widehat P_{\xi_1}}
\newcommand{\Pthm}[1]{\widehat P_{#1,m}}
\newcommand{\Fzero}{\mathcal F_0}
\newcommand{\Fone}{\mathcal F_1}
\newcommand{\xb}{\bar x}

\title{\bf Model Criticism via Stochastic Optimization}
\author{Ren-Yi Wang\\
Irving Institute for Cancer Dynamics, Columbia University, New York, NY, USA}
\date{\today}

\begin{document}
\maketitle

\begin{abstract}
Many generative models are specified by how data are simulated. An observation is a deterministic
transform $Y=G(\theta,X)$ of a draw $X$ from an assumed noise law, indexed by a structural parameter
$\theta$. Such models cover discrete, continuous and functional data, as well as i.i.d.\ replicates of
stochastic processes, but their likelihoods are often unavailable. We propose a simulation-based
procedure that tests whether such a model fits the data and, when it does not, locates \emph{where} it fails,
without choosing a test statistic. The procedure splits the data in two and is
formulated as a two-stage stochastic program. In the estimation stage, $\theta$ is fitted on the first
block by minimizing a Wasserstein distance between the data and simulations from the model. In the recourse stage, $\theta$ is frozen at the fitted $\tilde\theta$, and a finite set of representative
noise values $\{\bar x_j\}$ is pushed through the fitted mechanism to give model atoms
$G(\tilde\theta,\bar x_j)$. The held-out block optimizes weights $\bw$ on the
model atoms to best match its own distribution. The optimized weights are then compared with theoretical weights under correct specification. Since each model atom is the image of known noise values, the disagreement between optimal and theoretical weights points to the
part of the noise law that is wrong. The split makes the scaled weights conditionally
multinomial, yielding a calibrated Pearson test. We prove strong consistency of both stages under
checkable assumptions and illustrate the procedure on a Gaussian location--scale family, the number of
segregating sites under Kingman's coalescent, a Yule process, and an Ornstein--Uhlenbeck diffusion.
\end{abstract}

\section{Introduction}
\label{sec:introduction}

A generative model specifies a mechanism: an observation is a deterministic transform, indexed by a
structural parameter $\theta$, of a draw from an assumed noise law, $Y=G(\theta,X)$ with $X\sim\nu$.
The noise $X$ may live in any Polish space, such as a Euclidean space or a space of functions, and so
may the observation $Y$. The class therefore covers discrete and continuous observations, as
well as i.i.d.\ replicates of stochastic processes whose simulation algorithm is a deterministic map
of its driving randomness. For instance, continuous-time Markov chains are driven by Poisson random measures, and
diffusions by Brownian motion. Model criticism asks whether such a model is adequate for the data at
hand and, when it is not, \emph{where} it fails. For many generative models the likelihood is
unavailable or intractable, while simulation from $G$ is easy. We propose a criticism procedure that
uses only simulation and a distance between distributions, without the need to choose a test statistic. It is formulated as a two-stage stochastic program in the sense of \citet[Eq.~(1.9)]{PflugPichler2014}.

The procedure splits the observations into an estimation block $\xi_0$ and a held-out block $\xi_1$,
and uses them in two stages.

\emph{Estimation stage.} Using $\xi_0$, we fit the structural parameter $\theta$ while keeping the noise
law exactly as the modeler wrote it down. The estimate $\tilde\theta$ minimizes the $1$-Wasserstein
distance between the data and simulations from the model \citep{bassetti2006minimum,bernton2019parameter},
computed by sample average approximation \citep{shapiro2014lectures}. Minimum-distance estimators of this
kind are robust \citep{donoho1988automatic,bassetti2006asymptotic}, so a mild misfit of the noise law does not
drag $\tilde\theta$ far from the truth.

\emph{Recourse stage.} With $\theta$ frozen at $\tilde\theta$, we ask how the held-out data distribute
themselves relative to the fitted model. To make this question tractable, we choose a small set of
representative noise values $\xb_1,\dots,\xb_{N_1}$ (the \emph{mesh}), and push each through the fitted
mechanism $G(\tilde{\theta},\cdot)$. The resulting points $z_\ell=G(\tilde\theta,\xb_j)$ are the \emph{atoms} of the fitted model, interpreted as
typical observations the fitted model can produce, one for each representative noise value (noise values
that produce the same observation are merged into one atom). Each atom owns a \emph{cell}, the set of
observations closer to it than to any other atom. The fitted model assigns a probability to every cell and
the held-out block gives an empirical frequency for every cell. The recourse decision is the vector $\bw$
of weights on the atoms, and the optimal $\bw$ turns out to be exactly these held-out cell frequencies, in
closed form. Under a correct model the two sets of numbers agree; where they disagree, the model fails.
Because each atom is the image of known noise values, a cell with too much or too little held-out mass
points to the part of the noise distribution that is wrong. The cell counts also give a Pearson test of
the model, conditional on $\tilde{\theta}$.

The split is what makes this test calibrated, and splitting is familiar in Bayesian model criticism. In
\citet{moran2019population} and \citet{li2026calibrated} the split repairs a calibration defect as the
posterior predictive $p$-value is usually not uniform under the null \citep{gelman2013two}, and holding out
a block restores uniformity. There the two halves play the roles of ``train'' and ``test'' for one and the
same object, the predictive distribution. Our two halves instead determine two \emph{different} decisions.
$\xi_0$ determines $\theta$ with the noise law held at $\nu$, and $\xi_1$ determines $\bw$ with $\theta$ held at
$\tilde\theta$. Because $\tilde\theta$ is computed from $\xi_0$ alone, the held-out counts are, given
$\tilde\theta$, multinomial with known cell probabilities under the null. Hence, the test is calibrated without
any correction for estimation. Robustness of the first stage matters here as it guards $\tilde\theta$ against
mild misspecification of the structure $G$ or of the noise law $\nu$, so that the misfit is left for the
recourse stage to expose rather than absorbed into the fit.

Separating the estimation of a model from its criticism goes back to \citet{box1980sampling}, who proposed to estimate parameters through the posterior distribution and to criticize the model through the predictive distribution of the data, scoring the observed data against the prior predictive. \citet{gelman1996posterior} replaced the prior predictive by the posterior predictive distribution, so that the same data both fit and score the model. The resulting $p$-values are
not uniform under the null when the same data both fit and score the model \citep{bayarri2000pvalues}, and it motivates the split constructions of \citet{moran2019population} and
\citet{li2026calibrated} discussed above. Our structural stance is closest to \citet{covington2025bayesian},
who also criticize a model through a parametrization of its noise rather than through a statistic of the
observations.

The paper is organized as follows. Section~\ref{sec:prelim} recalls the Wasserstein distance and two-stage
stochastic programs, and casts model criticism as a two-stage program (Section~\ref{sec:two-stage}).
Section~\ref{sec:model} constructs the ingredients: the structural model, including stochastic processes, the
split, the synthetic noise of the estimation stage, and the recourse mesh. Section~\ref{sec:criticism} develops
the two stages, the prior test, the diagnostic and the calibrated test. Section~\ref{sec:assumptions} states the
assumptions under which both stages are consistent. Section~\ref{sec:experiments} illustrates the procedure on
four models, Section~\ref{sec:computation} describes the algorithms, and Section~\ref{sec:beyond} discusses
limitations and extensions. Proofs, and the verification of the assumptions for the examples are in the appendix.

\section{Preliminaries}
\label{sec:prelim}

\subsection{The space of measures and the Wasserstein distance}
\label{sec:wasserstein}

Throughout, the sample space $S$ is a Polish space, a separable metric space whose metric
$\dist$ is complete. The leading cases are the Euclidean space $S=\R^{s}$, which contains
integer-valued records such as counts, and path spaces such as
$L^1([0,T];\R^d)$, $L^2([0,T];\R^d)$ and $C([0,T];\R^d)$ with the supremum norm. When $S$ is a
normed space, $\dist(\omega_1,\omega_2)=\|\omega_1-\omega_2\|$. Let $\Psp(S)$ denote the set of
probability measures on $S$, and
\[
\PspOne(S) := \Big\{\mu \in \Psp(S) : \int_S \dist(\omega_0,\omega)\,\mu(d\omega) < \infty
\ \text{for some (all) } \omega_0 \in S\Big\}.
\]
For $\mu,\nu \in \PspOne(S)$ the
\emph{$1$-Wasserstein distance} is
\begin{equation}
\Wone(\mu,\nu)
:= \inf_{\pi \in \Pi(\mu,\nu)} \int_{S\times S} \dist(\omega_1,\omega_2)\,\pi(d\omega_1,d\omega_2),
\label{eq:W1-primal}
\end{equation}
the infimum being over the set of couplings $\Pi$ with marginals $\mu$ and $\nu$. By
Kantorovich--Rubinstein duality \citep[Ch.~2]{PflugPichler2014,Villani2009},
\begin{equation}
\Wone(\mu,\nu)
= \sup\Big\{\textstyle\int_S f\,d(\mu-\nu) : \|f\|_{\Lip}\le 1\Big\}.
\label{eq:W1-dual}
\end{equation}

The metric space $(\PspOne(S),\Wone)$ is complete and separable. Explicitly, if $C$
is a countable dense subset of $S$, then the set
of \emph{molecular probability measures} with rational coefficients on $C$,
\begin{equation}
\Big\{\textstyle\sum_{i=1}^{n} q_{i}\delta_{c_{i}}\ \Big|\ c_{i} \in C,\ 0 \leq q_{i} \in \mathbb{Q},\ \sum_{i=1}^{n}q_{i} = 1,\ n \in \N\Big\},
\label{eq:molecular-dense}
\end{equation}
is a countable dense subset of $(\PspOne(S),\Wone)$; see
\citet{dudley2018real} and \citet{dull2021spaces} for more details. In contrast, $\PspOne(S)$ equipped with $\|\cdot\|_{TV}$ is not separable as shown in Example~\ref{ex:diracs}. In addition, $\Wone$ metrizes weak convergence together with convergence of first
moments on $\PspOne(S)$ \citep[Thm.~6.9]{Villani2009}.

\begin{example}\label{ex:diracs}
For $a\ne b$, $\|\delta_a-\delta_b\|_{\TV} = 1$, so $a_{n}\to b$ in
$S$ does \emph{not} imply
$\|\delta_{a_{n}}-\delta_{b}\|_{\TV}\to0$. By contrast
$\Wone(\delta_{a_{n}},\delta_{b})=\dist(a_{n},b)$ and $a_{n}\to b$ implies $\Wone(\delta_{a_{n}},\delta_{b}) \to 0$.
\end{example}

\subsection{Two-stage stochastic programs}
\label{sec:sto-primer}

A stochastic optimization problem selects a decision to optimize a
cost that depends on uncertainty whose distribution is known. In the
canonical form of \citet[Eq.~(1.7)]{PflugPichler2014},
\begin{equation}
\min_{x\in\mathbb X}\ R_P\big[Q(x,\xi)\big],
\label{eq:sp-generic}
\end{equation}
$\xi\sim P$ is the random datum, $x\in\mathbb X$ the decision, $Q(x,\xi)$
the resulting cost, and $R_P$ a \emph{risk functional} summarizing the random
cost by a scalar. Taking $R_P=\E_P$ gives a \emph{risk-neutral} program.
When $P$ is not tractable, the \emph{sample average approximation} (SAA)
replaces it by the empirical measure of scenarios drawn once and for all
from $P$ and solves the resulting deterministic program
\citep[Ch.~5]{shapiro2014lectures}.

Decisions are random variables constrained to use only the information
available when they are taken. A filtration
$\Fzero\subseteq\Fone\subseteq\cdots$ encodes that flow, and a decision
fixed at stage $t$ must be $\mathcal F_t$-measurable. The two extremes are a \emph{here-and-now}
decision, fixed before the relevant uncertainty is observed, and a
\emph{wait-and-see} decision, taken after the realization and so free to
depend on it. A two-stage \emph{recourse} program fixes $x_0$ here-and-now,
observes $\xi_1$, and then takes a recourse (corrective) action in the feasible set
$x_1\in\mathbb X_1(x_0,\xi_1)$. Section~\ref{sec:two-stage} casts model criticism in this form.

\subsection{Model criticism as a two-stage program}
\label{sec:two-stage}

We ask two questions in sequence. First, which structure within the model class describes the data
best, when the noise law is taken exactly as the modeler wrote it down? Second, with that structure fixed, how do
\emph{new} data redistribute the mass that the fitted model assigns to its own support? We answer the two questions
by the two stages of a stochastic program on a sample split $(\xi_0,\xi_1)$, with the following correspondence:

\begin{center}
\small
\begin{tabular}{p{2.0cm}p{6.0cm}p{6.6cm}}
\hline
 & Stage 0 (here-and-now) & Stage 1 (recourse)\\
\hline
Decision & Structural parameter $\theta\in K\subseteq\Theta$ & Weights $\bw\in\Delta^{L-1}$ on the atoms of $P_{\tilde\theta}$\\
Information & Estimation block $\xi_0$ and synthetic noise & Held-out block $\xi_1$, with $\tilde\theta$ fixed\\
Cost & $\Wone\big(\Pxo,P_\theta\big)$ & $\Wone\big(\Pxi,P_{\tilde\theta,\bw}\big)$\\
Interpretation & Best structure under the assumed noise law & How the held-out data reweight the fitted model\\
\hline
\end{tabular}
\end{center}

\noindent Written out, the program is
\begin{equation}
\boxed{\ \
\tilde\theta\ \in\ \argmin_{\theta\in K}\ \Wone\big(\Pxo,\,P_\theta\big),
\qquad
\tilde\bw\ \in\ \argmin_{\bw\in\Delta^{L(\tilde\theta)-1}}\ \Wone\Big(\Pxi,\ \textstyle\sum_{\ell}w_\ell\,\delta_{z_\ell(\tilde\theta)}\Big),
\ \ }
\label{eq:two-stage-overview}
\end{equation}
where $z_1(\tilde\theta),\dots,z_{L(\tilde\theta)}(\tilde\theta)$ are atoms of the fitted model obtained by pushing a finite
mesh of the noise space through $G(\tilde\theta,\cdot)$ (Section~\ref{sec:mol_approx}). The first stage optimizes over the
model class and the second optimizes \emph{within} the fitted model $P_{\tilde\theta}$. The recourse decision reweights the fitted model's own atoms, chosen after $\xi_1$ has been observed. Under correct specification the optimal
reweighting returns the masses that the fitted model itself assigns to the cells of its atoms, $w^{\mathrm{vor}}(\tilde\theta)$.
The departure between optimized $\tilde\bw$ and theoretical $w^{\mathrm{vor}}(\tilde\theta)$ is the diagnostic, and its conditionally multinomial law gives a calibrated
test. In the language of Section~\ref{sec:sto-primer}, $\theta$ is fixed before $\xi_1$ is revealed and is
$\Fzero$-measurable, while $\bw$ is a wait-and-see decision that may depend on $\xi_1$.

We note a difference from a classical recourse program. In the classical setting, the first-stage decision
anticipates the recourse and minimizes the expected total cost,
\[
Q_0(x_0)+\E\big[\min_{x_1}Q_1(x_0,x_1;\xi_1)\big].
\]
Here
the first stage minimizes its own cost alone. The program is \emph{sequential}, and the first stage is myopic by design.
Were $\theta$ chosen to anticipate the recourse cost, it would bend to absorb a misfit of the noise law, and the departure
that the recourse is meant to expose would disappear into $\tilde\theta$. Section~\ref{sec:criticism} develops both stages,
and Figure~\ref{fig:schematic} summarizes the procedure.

\begin{figure}[t]
\centering
\resizebox{\textwidth}{!}{%
\begin{tikzpicture}[font=\small, >=Latex,
  box/.style={draw, rounded corners=2pt, align=center, inner sep=4pt, minimum height=1.0cm}]
\node[box, fill=blue!5]   (data0) at (0,0)    {estimation block $\xi_0$\\ synthetic noise $\tilde x_{1:m}\sim\nu$};
\node[box, fill=blue!12]  (s0)    at (5.2,0)  {\textbf{Stage 0} (here-and-now)\\ $\tilde\theta\in\argmin_{\theta\in K}\Wone\big(\Pxo,\Pthm{\theta}\big)$};
\node[box, fill=orange!12](s1)    at (11.2,0) {\textbf{Stage 1} (recourse)\\ $\tilde\bw\in\argmin_{\bw}\Wone\big(\Pxi,P_{\tilde\theta,\bw}\big)$};
\node[box, fill=green!6]  (mesh)  at (9.2,2.3){mesh $\xb_{1:N_1}$: $N_1$-medoid\\ of simulated noise};
\node[box, fill=orange!5] (data1) at (13.3,2.3){held-out block $\xi_1$};
\node[box, fill=gray!10]  (prior) at (5.2,-2.1){prior test:\\ is $\tilde\theta$ typical under $\Pi$?};
\node[box, fill=gray!10]  (out)   at (11.2,-2.1){profile $\tilde\bw$ against $w^{\mathrm{vor}}(\tilde\theta)$\\ and calibrated Pearson test};
\draw[->, thick] (data0) -- (s0);
\draw[->, thick] (s0) -- node[above]{$\tilde\theta$} (s1);
\draw[->, thick] (data1.south) -- (data1.south |- s1.north);
\draw[->, thick] (mesh.south) -- node[left]{atoms $G(\tilde\theta,\xb_j)$} (mesh.south |- s1.north);
\draw[->, thick] (s0) -- (prior);
\draw[->, thick] (s1) -- (out);
\begin{scope}[shift={(0,-6.2)}]
  \draw[->] (-2.6,0) -- (2.8,0) node[right]{$\Xs$};
  \draw[blue!60!black, thick, domain=-2.4:2.4, samples=80] plot (\x, {1.5*exp(-\x*\x/1.2)});
  \node[blue!60!black] at (-1.9,1.2) {$\nu$};
  \foreach \x in {-2.0,-1.35,-0.85,-0.42,0,0.42,0.85,1.35,2.0}
     \fill[black] (\x,0) circle (2pt);
  \foreach \x in {-1.675,-1.1,-0.635,-0.21,0.21,0.635,1.1,1.675}
     \draw[gray] (\x,-0.12) -- (\x,0.12);
  \node[align=center] at (0.1,-0.75) {noise space: mesh points $\xb_j$ (dots) and cells $C_j$};
  \draw[->, very thick, gray] (3.3,0.6) -- node[above]{$G(\tilde\theta,\cdot)$} (5.6,0.6);
  \begin{scope}[shift={(9.6,0)}]
    \draw[->] (-3.6,0) -- (3.9,0) node[right]{$S$};
    \foreach \z/\h/\m/\c in {-3.0/0.35/0.40/orange!70, -2.05/0.75/0.80/orange!70, -1.3/1.05/1.10/orange!70,
                             -0.62/1.25/1.30/orange!70, 0/1.35/1.35/orange!70, 0.62/1.30/1.30/orange!70,
                             1.3/0.55/1.10/blue!70!black, 2.05/0.85/0.80/orange!70, 3.0/1.45/0.40/red!80!black}
       {\fill[\c] (\z-0.18,0) rectangle (\z+0.18,\h);
        \draw[very thick] (\z-0.24,\m) -- (\z+0.24,\m);
        \fill[black] (\z,0) circle (1.6pt);}
    \node[red!80!black, align=center] at (3.0,2.0) {excess};
    \node[blue!70!black, align=center] at (1.3,1.75) {deficit};
    \node[align=center] at (0.1,-0.75) {observation space: atoms $z_\ell(\tilde\theta)$, held-out weights $\tilde w_\ell$ (bars),\\ model masses $w^{\mathrm{vor}}_\ell(\tilde\theta)$ (ticks)};
  \end{scope}
\end{scope}
\end{tikzpicture}}
\caption{The procedure at a glance. Top: the estimation block and the synthetic noise determine $\tilde\theta$ (Stage 0),
which is scored against the prior; the mesh, a finite set of simulated noise values, is pushed through $G(\tilde\theta,\cdot)$
to give the atoms of the fitted model, and the held-out block reweights them (Stage 1). Bottom: an illustration for a
one-dimensional Gaussian location--scale model. The held-out weights $\tilde w_\ell$ are compared atom by atom with the model's
masses; a cell above its pointwise interval is reported as excess (red), a cell below it as deficit (blue).}
\label{fig:schematic}
\end{figure}

\section{The structural model, the split, synthetic noise, and the recourse mesh}
\label{sec:model}

\subsection{Mechanism}
\label{sec:mechanism}

Following \citet{covington2025bayesian}, we posit that the data-generating
mechanism admits a structural representation. The structural parameter $\theta$ is treated as a draw from a prior $\Pi$. The parameter set $\Theta$ is a set equipped with a metric $\rho$ such that
\begin{equation}
\Pi(\Theta)=1,
\qquad \theta_0\in\Theta .
\label{eq:prior-support}
\end{equation}
Throughout, $\|\theta-\theta'\|$ denotes $\rho(\theta,\theta')$ and in $\Theta\subseteq\R^{p}$ $\rho$ is the Euclidean distance. An observation is a deterministic transform, indexed by the structural parameter $\theta$, of a draw from
a noise law $\nu$. Let the \emph{noise space} $\Xs$ be a Polish space, let $\nu\in\Psp(\Xs)$ be any
Borel probability measure on it, and let
\begin{equation}
X_1,\dots,X_n\ \stackrel{\text{i.i.d.}}{\sim}\ \nu,
\qquad
Y_t\;=\;G\big(\theta_0,X_t\big)\in S,\qquad t=1,\dots,n.
\label{eq:structural}
\end{equation}
The procedure uses $\nu$ only through simulation. In the estimation stage, we simulate a synthetic sample
(Section~\ref{sec:synthetic}), and compute a finite mesh of the noise space from simulated draws in the recourse stage (Section~\ref{sec:mol_approx}). The mechanism
$G:\Theta\times\Xs\to S$ is Borel measurable.

\subsection{Stochastic processes as mechanisms}
\label{sec:processes}

For a stochastic process,
an observation $Y_t$ is a whole replicate of the process, recorded either at finitely many times
$0<t_1<\dots<t_J\le T$, so that $S\subseteq\R^{dJ}$, or as a path on $[0,T]$, so that $S$ is a path
space. The replicates $Y_1,\dots,Y_n$ must be independent, as for independent lineages, cells,
or patients. 

\paragraph{Continuous-time Markov chains.}
Let $Z$ take values in $E\subseteq\mathbb Z^d$ and jump by $\zeta_r\in\mathbb Z^d$ at rate
$\lambda_r(\theta,Z)\ge0$, $r=1,\dots,R$. Then $Z$ is the solution of the stochastic differential equation
driven by Poisson random measures
\begin{equation}
Z(t)\ =\ z_0+\sum_{r=1}^{R}\zeta_r\int_0^t\!\!\int_0^\infty\mathbf 1\big\{u\le\lambda_r\big(\theta,Z(s-)\big)\big\}\,N_r(ds,du),
\qquad t\in[0,T],
\label{eq:poisson-sde}
\end{equation}
where $N_1,\dots,N_R$ are independent Poisson random measures on $[0,T]\times[0,\infty)$ with intensity
$ds\,du$ \citep[Ch.~II]{ikeda1989stochastic,fournier2004microscopic}. Each atom $(s,u)$ of $N_r$ is a candidate
jump of type $r$ at time $s$, and it is accepted when $u\le\lambda_r(\theta,Z(s-))$, interpreted as Poisson thinning. The noise is $x=(N_1,\dots,N_R)$, the sample paths of the Poisson random measures. Birth--death processes correspond to $d=1$, $\zeta_1=+1$, $\zeta_2=-1$ \citep{kendall1948generalized} and stochastic chemical reaction networks are the general case \citep{gillespie1977exact,anderson2015stochastic}. Paths can be
recorded in $L^1([0,T];\R^d)$, a separable Banach space. The Skorokhod space of c\`adl\`ag paths is also Polish
\citep[Sec.~12]{billingsley1999convergence}.

\paragraph{Diffusions.}
Let $Z$ solve $dZ(t)=b(\theta,Z(t))\,dt+\sigma(\theta,Z(t))\,dW(t)$, $Z(0)=z_0\in\R^d$, with $W$ an
$m$-dimensional Brownian motion. The noise space is $\Xs=C_0([0,T];\R^m)$, the continuous paths
started at $0$ with the supremum norm, and $\nu$ is Wiener measure. Observations are snapshots in
$\R^{dJ}$ or paths in $C([0,T];\R^d)$ or $L^2([0,T];\R^d)$.

\subsection{The split}
\label{sec:split}

Partition the $n$ observations into
\begin{equation}
\xi_0=(y_1,\dots,y_{n_0}),\qquad
\xi_1=(y_{n_0+1},\dots,y_{n}),\qquad n_0+n_1=n,
\label{eq:split}
\end{equation}
with empirical measures
$\Pxo=\frac1{n_0}\sum_{t\le n_0}\delta_{y_t}$ and
$\Pxi=\frac1{n_1}\sum_{t>n_0}\delta_{y_t}$. The block sizes $n_0=n_0(n)$ and $n_1=n_1(n)$ are
indexed by the total sample size $n$, and throughout we require
\begin{equation}
n_0\wedge n_1\ \longrightarrow\ \infty,
\qquad
\frac{n_1}{\log n}\ \longrightarrow\ \infty,
\label{eq:regime}
\end{equation}
the second condition holding for instance when $n_1$ is a fixed fraction of $n$.

\subsection{Synthetic noise for the estimation stage}
\label{sec:synthetic}

The model law $P_\theta$ is available through simulation. Fix a synthetic
sample size $m=m(n_0)\in\N$ and draw,
\begin{equation}
\tilde x_1,\dots,\tilde x_m\ \stackrel{\text{i.i.d.}}{\sim}\ \nu,
\qquad
\Pthm{\theta}\ :=\ \frac1m\sum_{i=1}^{m}\delta_{G(\theta,\tilde x_i)} .
\label{eq:synthetic}
\end{equation}
The same noise draws are used for every $\theta$. The empirical measure $\Pthm{\theta}$ is the \emph{sample average approximation} (SAA) of $P_\theta$
\citep[Ch.~5]{shapiro2014lectures}, and it plays the role of the
\emph{scenario generation} step of \citet{PflugPichler2014}. It is also the
single-replicate Monte Carlo approximation of the minimum expected
Wasserstein estimator used in the computations of
\citet[Sec.~3.2]{bernton2019parameter}, where the noise is likewise drawn once
and $\theta$ is optimized with the draws held fixed.

Together with the split, the synthetic sample forms the filtration
\begin{equation}
\Fzero=\sigma(\xi_0,\tilde x_{1:m})\ \subset\ \Fone=\sigma(\xi_0,\tilde x_{1:m},\xi_1).
\label{eq:filtration}
\end{equation}
Because $\tilde x_{1:m}$ is independent of the data, conditioning on $\Fzero$
leaves $\xi_1$ an i.i.d.\ sample from the law that generated the data.

\subsection{The recourse mesh: the $N_1$-medoid of simulated noise}
\label{sec:mol_approx}

The recourse stage needs a fixed, finite set of model atoms at which a departure from the model
can be reported. We obtain them by pushing finitely many noise values through the fitted
mechanism, and we choose these noise values to represent $\nu$ as well as $N_1$ points can, in the
sense of optimal quantization \citep{graf2000foundations}.

\paragraph{Optimal quantization.}
Fix a \emph{quantization (pseudo)metric} $\dist_Q$ on $\Xs$, Borel measurable on $\Xs\times\Xs$. Fix an exponent $r\ge1$. For a mesh
$\Gamma=\{\xb_1,\dots,\xb_{N_1}\}\subset\Xs$ let
\begin{equation}
C_j:=\Big\{x\in\Xs:\ j=\argmin_{j'}\dist_Q(x,\xb_{j'})\Big\},
\qquad\text{ties broken by lowest index},
\label{eq:noise-cells}
\end{equation}
be its Voronoi cells, $q_j:=\nu(C_j)$ their masses, and
\begin{equation}
e_r(\Gamma;\nu)\ :=\ \Big(\int_{\Xs}\min_{j\le N_1}\dist_Q(x,\xb_j)^r\,\nu(dx)\Big)^{1/r}
\label{eq:distortion}
\end{equation}
its quantization error. Optimal quantizers exist for $\Xs=\R^k$ when $d_{Q}$ is a metric and 
\[
\int\dist_Q(x_0,x)^r\,\nu(dx)<\infty
\]
\citep{graf2000foundations,luschgy2002functional}. In a general Polish space an optimal quantizer need not exist.

\paragraph{Construction from simulated noise.}
Because $\nu$ is available through simulation, the mesh is computed from simulated noise
\begin{equation}
x^{(1)},\dots,x^{(M)}\ \stackrel{\text{i.i.d.}}{\sim}\ \nu,
\label{eq:quant-sample}
\end{equation}
which can also be subsampled from the $m$ synthetic samples in previous stage with $M < m$. We take the mesh to be their \emph{$N_1$-medoid}, that is,
\begin{equation}
\widehat\Gamma\ \in\ \argmin\Big\{\,\textstyle\frac1M\sum_{i\le M}\min_{\bar x\in\Gamma}\dist_Q\big(x^{(i)},\bar x\big)^r:\
\Gamma\subseteq\{x^{(1)},\dots,x^{(M)}\},\ |\Gamma|=N_1\Big\},
\label{eq:medoid}
\end{equation}
computed approximately by Algorithm~\ref{alg:quantize} of Section~\ref{sec:quantize}. Every mesh point is a genuine draw from $\nu$, so $G(\theta,\xb_j)$ is
always defined and typical of the model.

\paragraph{Raw and merged atoms.}
The \emph{raw atoms} are $G(\theta,\xb_j)$, $j=1,\dots,N_1$. When $G$ is
many-to-one, distinct mesh points can have the same image.
We therefore merge coinciding images. Let
$z_1(\theta),\dots,z_{L(\theta)}(\theta)$ be the \emph{distinct} values among
$\{G(\theta,\xb_j)\}_{j\le N_1}$, and let
\begin{equation}
I_\ell(\theta)\ :=\ \big\{j\le N_1:\ G(\theta,\xb_j)=z_\ell(\theta)\big\},
\qquad \ell=1,\dots,L(\theta),
\label{eq:fibers}
\end{equation}
be the \emph{fibers}, labelled so that $\min I_1(\theta)<\min I_2(\theta)<\cdots$.
The fibers partition the mesh and the \emph{molecular model measure} on the merged atoms is
\begin{equation}
P_{\theta,\bw}\ :=\ \sum_{\ell=1}^{L(\theta)} w_\ell\,\delta_{z_\ell(\theta)},
\qquad \bw\in\Delta^{L(\theta)-1},
\label{eq:atoms}
\end{equation}
where $\Delta^{L-1}=\{\bw\in[0,1]^L:\sum_\ell w_\ell=1\}$. The weights $\bw$ are
the \emph{model's weights}, living on observation-space atoms. When $G$ is continuous and injective in $x$ the fibers are singletons,
$L(\theta)=N_1$. When $P_\theta$ is finitely supported, $L(\theta)$
is at most the size of the support, however fine the mesh.

\section{Model criticism procedure}
\label{sec:criticism}

The sample split \eqref{eq:split} splits the decision as well. The structural
parameter is chosen first with the noise law held at its assumed value, and
the weights are chosen second with the structure held at its fitted value. The
here-and-now decision is therefore the structural parameter $\theta$ alone and the
recourse decision is the weight vector $\bw$ alone,
\begin{equation}
\boxed{\ \
\mathbb X_0(\xi_0)\ :=\ K\subseteq\Theta,
\qquad
\mathbb X_1\big(\tilde\theta,\xi_1\big)\ :=\ \Delta^{L(\tilde\theta)-1},
\ \ }
\label{eq:feasible}
\end{equation}
with $K$ the compact search set of Assumption~\ref{ass:search}.
The stage costs are Wasserstein discrepancies against the two blocks,
\begin{equation}
Q_0\big(x_0;\xi_0\big):=\Wone\big(\Pxo,\,\Pthm{\theta}\big),
\qquad
Q_1\big(x_0,x_1;\xi_1\big):=\Wone\big(\Pxi,\,P_{\tilde\theta,\bw}\big),
\label{eq:stagecosts}
\end{equation}
the first formed against the synthetic sample \eqref{eq:synthetic} and the second on the merged atoms \eqref{eq:atoms}. Ideally, we would like to compute $\Wone\big(\Pxo,\,P_{\theta}\big)$ exactly in the first stage, but this is not feasible for general $P_{\theta}$; we therefore replace $P_\theta$ by its SAA $\Pthm{\theta}$, and no mesh is used at this stage.

We explain the design of the feasible sets \eqref{eq:feasible}. First, the estimation stage may not touch the weights. Simulating from $\nu$ itself indicates that when the structure is fitted, the noise law is
\emph{assumed} instead of estimated. Second, the recourse stage may not revise the structure. Were $\theta$ free at the recourse stage, any noise-law departure could be absorbed by refitting the structure, rendering the diagnostic uninterpretable.

\subsection{Estimation stage}
Let $\theta_0$ be a draw from $\Pi$
generating $y_{1:n}$ through the mechanism \eqref{eq:structural}. We solve the estimation-stage term alone,
\begin{equation}
\thn\ \in\ \varepsilon_{n_0}\text{-}\argmin_{\theta\in K}\ \Wone\big(\Pxo,\,\Pthm{\theta}\big),
\qquad m=m(n_0),
\label{eq:stage0}
\end{equation}
where $\varepsilon\text{-}\argmin_{\theta} f:=\{\theta: f(\theta)\le\varepsilon+\inf f\}$ and
$\varepsilon_{n_0}\downarrow0$. We call $\thn$ the \emph{SAA minimum Wasserstein estimator}.

When the observations are vector-valued, $S=\R^s$ with $s>1$, every evaluation of the objective in \eqref{eq:stage0} is a
transport linear program between $n_0$ and $m$ points (Section~\ref{sec:W1eval}), and the search of
Algorithm~\ref{alg:estimation} becomes inefficient. For efficiency we then replace $\Wone$ at the estimation
stage by the sliced Wasserstein distance, the average of one-dimensional $\Wone$ distances between the projections of the
two samples on a fixed set of directions \citep{nadjahi2019asymptotic}. This changes
the estimator, whose consistency is then that of minimum sliced-Wasserstein estimators \citep{nadjahi2019asymptotic}
rather than Theorem~\ref{thm:consistency}.

For large $n_0$
a consistent estimate behaves like the draw $\theta_0$ itself, and should therefore look like a typical draw from $\Pi$. To test if $\theta$ is a typical value for the prior $\Pi$, we set up the hypothesis with significance level $\alpha = 0.05$:
\begin{align*}
    H_0:\theta\sim\Pi \text{ against } H_1=\neg H_0.
\end{align*}
Since we observe only one instantiation of the structural parameter $\theta_0$ from the prior, indirectly through the data $Y_{1},\cdots,Y_{n}$, detection of prior misspecification has low power, regardless of the chosen test. We take a summary statistic $T(\theta)$, by default the Euclidean norm $T(\theta)=\|\theta\|$, write $F(t):=\Pi(T(\theta)\le t)$ and $F(t^-):=\Pi(T(\theta)<t)$, and set
\begin{align*}
    U = F(t^-) + V\,\Pi\big(T(\theta)=t\big),\qquad t=T(\tilde{\theta}_{n_{0}}),\qquad V\sim\mathrm{Unif}(0,1)\ \text{independent of the data},
\end{align*}
and the $p$-value to be
\begin{align*}
    p = 2\min\{U,\ 1-U\}.
\end{align*}

The central estimation-stage result is the strong consistency of $\thn$
(Theorem~\ref{thm:consistency}, Appendix~\ref{sec:proofs}). Under
Assumptions~\ref{ass:modulus}--\ref{ass:search}, for any measurable
$\varepsilon_{n_0}$-approximate minimizer $\thn$ of Equation~\eqref{eq:stage0} with
$\varepsilon_{n_0}\downarrow0$, and if the synthetic sample grows with the estimation block,
$m=m(n_0)\to\infty$,
\begin{align*}
    \lim_{n_{0}\to\infty}\|\thn-\theta_0\|= 0 \text{ almost surely.}
\end{align*}
Hence, $\thn$ concentrates near $\theta_0$ for large $n_{0}$, implying $\thn$ should be a typical draw from the prior $\Pi$. Deviation suggests either the prior or the noise law may be misspecified. Since the structural parameter, the noise law, and the structural equation are all entangled, we assume misspecification in the structural equation $G$ can be absorbed in the noise law and when we identify misspecification in the noise law, we may also attribute it to the structure $G$, depending on the context (see the phylogenetic example in Section \ref{sec:exp-coal}).

To protect the influence of noise law misspecification on the structural parameter. We show the estimator $\tilde{\theta}_{n_{0}}$ is qualitatively robust. Define the estimation functional $T$ on $(\mathcal{P}_{1}(S), \Wone)$ by $T(Q)\in\argmin_{\theta\in K}\Wone(Q,P_\theta)$, so that $\thn$ is an approximation of $T(\Pxo)$; we say $T$ is qualitatively robust at the model if it is continuous at $P_{\theta_0}$ \citep{hampel1971general}. In Proposition~\ref{prop:robust} we show that if $Q_k\to P_{\theta_0}$ in $\Wone$, then every sequence of approximate minimizers of $\theta\mapsto\Wone(Q_k,P_\theta)$ over $K$, and likewise of its SAA version, converges to $\theta_0$. This yields \emph{qualitative}
robustness \citep{ronchetti2009robust}, namely continuity of $T$ at the model on $(\PspOne(S),\Wone)$,
and stability under small $\Wone$ mismatch; related robustness properties of minimum-dissimilarity estimators are established by \citet{bassetti2006asymptotic}, and illustrated numerically under misspecification and heavy tails by \citet[Secs.~4.3--4.4]{bernton2019parameter}.

\subsection{Recourse stage}
\label{sec:closedform}

Fix $\tilde\theta:=\thn$ from the previous stage and write $z_\ell:=z_\ell(\tilde\theta)$,
$\ell=1,\dots,L:=L(\tilde\theta)$, for the merged atoms of Section~\ref{sec:mol_approx}. The
recourse-stage problem is
\begin{equation}
Q(\tilde\theta)
=\min_{\bw\in\Delta^{L-1}}\Wone\big(\Pxi,P_{\tilde\theta,\bw}\big).
\label{eq:stage1-primal}
\end{equation}
The recourse problem is solved by \emph{Voronoi recourse} \citep[Lem. 4.9]{PflugPichler2014}. The minimizer
and the value of Equation~\eqref{eq:stage1-primal} are
\begin{equation}
\tilde w_\ell
=\frac1{n_1}\#\big\{y\in\xi_1:\ \ell=\argmin_{\ell'}\dist(y,z_{\ell'})\big\},
\qquad
Q(\tilde\theta)=\frac1{n_1}\sum_{y\in\xi_1}\min_{\ell}\dist\big(y,z_{\ell}\big),
\label{eq:voronoi}
\end{equation}
ties broken by lowest label. Define the observation-space Voronoi cells of the merged atoms and the
\emph{model's Voronoi masses}
\begin{equation}
w^{\mathrm{vor}}_\ell(\theta):=P_{\theta}\big(V_\ell(\theta)\big),
\qquad
V_\ell(\theta):=\big\{y\in S:\ \ell=\argmin_{\ell'}\dist\big(y,z_{\ell'}(\theta)\big)\big\},
\label{eq:wvor}
\end{equation}
ties broken by lowest label. If $P_\theta$ is purely atomic and every atom of
$P_\theta$ is a merged atom, each cell contains exactly one atom of
$P_\theta$ and
\begin{equation}
w^{\mathrm{vor}}_\ell(\theta)=P_\theta\big(\{z_\ell(\theta)\}\big)
=\nu\big\{x\in\Xs:\ G(\theta,x)=z_\ell(\theta)\big\},
\label{eq:wvor-discrete}
\end{equation}
the probability mass function of the model at the atom. In that case
\eqref{eq:voronoi} is the relative frequency of each support value in the
held-out block, and held-out values outside the model's support are
counted into the nearest atom, typically an extreme one.

\begin{remark}[Identification]\label{rmk:fiber}
At the resolution of the mesh, the identified object is the vector of fiber totals
$\tilde w_\ell$, $\ell\le L$; how the mass is distributed \emph{within} a fiber
$I_\ell(\tilde\theta)$ is not identified. When a noise-space picture is
wanted, we display on the noise cells $C_j$ of \eqref{eq:noise-cells} the $\nu$-proportional lift
$\tilde w_\ell\, q_j/\sum_{j'\in I_\ell}q_{j'}$ for
$j\in I_\ell$. This is a display convention, not a rigorous estimate.
\end{remark}

At the recourse stage we localize the residual misspecification in the noise
law. In Theorem~\ref{thm:stage1-weights}, Appendix~\ref{sec:proofs}, we show that almost surely, as $n_0,n_1\to\infty$, the number of merged atoms eventually equals $L(\theta_0)$ and, for each $\ell$,
\begin{align*}
    \tilde w_\ell\to w^{\mathrm{vor}}_\ell(\theta_0),\qquad
    w^{\mathrm{vor}}_\ell(\thn)\to w^{\mathrm{vor}}_\ell(\theta_0),\qquad
    \sum_\ell\tilde w_\ell\delta_{z_\ell(\thn)}\ \xrightarrow{\ \Wone\ }\ \sum_\ell w^{\mathrm{vor}}_\ell(\theta_0)\delta_{z_\ell(\theta_0)} .
\end{align*}

Condition on $\Fzero$, which fixes $\thn$, and hence fixes the merged atoms and the cells $V_\ell(\thn)$. This leaves the held-out
observations i.i.d.\ from the law that actually generated them, and write
\begin{equation}
p_\ell\ :=\ \Prob\big\{Y\in V_\ell(\thn)\ \big|\ \Fzero\big\},
\qquad \ell=1,\dots,L,
\label{eq:cellprob}
\end{equation}
for the resulting cell probabilities. Since the cells partition $S$ we have
$\sum_\ell p_\ell=1$, and Equation~\eqref{eq:voronoi} counts each held-out observation into
exactly one of them, so by construction
\begin{equation}
\big(n_1\tilde w_1,\dots,n_1\tilde w_{L}\big)\ \big|\ \Fzero
\ \sim\ \mathrm{Multinomial}\big(n_1;\,p_1,\dots,p_{L}\big).
\label{eq:multinomial}
\end{equation}
The counts are not independent across cells as they sum to $n_1$ by construction, and
\begin{equation}
\mathrm{sd}\big(\tilde w_\ell\mid\Fzero\big)=\sqrt{\frac{p_\ell(1-p_\ell)}{n_1}},
\qquad
\mathrm{corr}\big(\tilde w_\ell,\tilde w_{\ell'}\mid\Fzero\big)=-\sqrt{\frac{p_\ell\,p_{\ell'}}{(1-p_\ell)(1-p_{\ell'})}}.
\label{eq:cellsd}
\end{equation}
Replacing $p_\ell$ by its value under the fitted model, $w^{\mathrm{vor}}_\ell(\thn)$, gives the
per-cell error bar for the diagnostic. With continuous observations and singleton fibers the masses are of order
$1/N_1$ and the correlations are $O(N_1^{-1})$. With discrete observations the masses are the
model's probability mass function \eqref{eq:wvor-discrete} and need not be small, so the
correlations should be kept in mind when a few atoms carry most of the mass. The signal-to-noise ratio of cell $\ell$ is
\begin{align*}
    \frac{w^{\mathrm{vor}}_\ell(\thn)}{\mathrm{sd}\big(\tilde w_\ell\mid\Fzero\big)} \approx \sqrt{\frac{n_{1}\,w^{\mathrm{vor}}_\ell(\thn)}{1-w^{\mathrm{vor}}_\ell(\thn)}}\ \ge\ \sqrt{e_\ell},
    \qquad e_\ell:=n_1\,w^{\mathrm{vor}}_\ell(\thn),
\end{align*}
the square root of the expected count. In the continuous case $e_\ell$ is of order $n_1/N_1$, so refining the recourse mesh trades a finer resolution for a noisier estimate of every individual weight. In the discrete case the resolution is capped by the support and refining the mesh beyond it has no effect on the atoms. A practical default is to choose $N_1$ so that every expected count $e_\ell$ is at least in the tens, which is also suggested by \citet[Alg. 4.2]{PflugPichler2014} for optimal quantization. This also guarantees the signal-to-noise ratio is at least $3$, which satisfies the limit of detection standard in \cite{borman2017q2}.

The natural test compares the observed counts in Equation~\eqref{eq:multinomial}
with the counts the fitted model predicts. Conditionally on $\Fzero$, we hypothesize
\begin{align}
H_0:\ p_\ell = w^{\mathrm{vor}}_\ell(\thn)\ \text{ for every } \ell=1,\dots,L,
\qquad
H_1:\ p_\ell \neq w^{\mathrm{vor}}_\ell(\thn)\ \text{ for some } \ell.
\label{eq:chisq-hyp}
\end{align}
$H_0$ asserts that the fitted model reproduces the held-out cell
probabilities. The Pearson statistic is
\begin{equation}
T_{n_1}\ :=\ \sum_{\ell=1}^{L}
\frac{\big(n_1\tilde w_\ell-n_1w^{\mathrm{vor}}_\ell(\thn)\big)^2}{n_1w^{\mathrm{vor}}_\ell(\thn)},
\qquad
T_{n_1}\mid\Fzero\ \Longrightarrow\ \chi^2_{L-1}
\label{eq:chisq}
\end{equation}
under $H_0$ as $n_1\to\infty$, and we reject when $T_{n_1}$ exceeds the $(1-\alpha)$ quantile of
$\chi^2_{L-1}$. The test is \emph{calibrated by the split} as $\thn$ is
$\Fzero$-measurable and $\xi_1$ is independent of $\Fzero$. Following the procedure in \citep{cochran1954some}, adjacent cells are pooled until every expected count
$e_\ell$ is at least $5$, the degrees of freedom being reduced to the number of pooled
cells minus one.

If we reject the null hypothesis, the optimized recourse weight profile $\tilde{\mathbf{w}}$ provides a visual
diagnostic for which aspects of the noise law or structure are misspecified. The optimized $\tilde{\mathbf{w}}$ is compared atom by atom with the reference weight $w^{\mathrm{vor}}(\thn)$, with per-cell error bar
\eqref{eq:cellsd} widened from one standard deviation to $1-\alpha$
coverage. When the correlations in \eqref{eq:cellsd} are negligible, we may treat the bars as bona fide $100(1-\alpha)\%$ confidence intervals. The discrepancy between $\tilde{\mathbf{w}}$ and $w^{\mathrm{vor}}(\thn)$ says which part of the noise distribution is wrong; by Remark~\ref{rmk:fiber} it does so at the level of the fibers of the fitted mechanism. 

\section{Assumptions}
\label{sec:assumptions}

In this section we collect assumptions that are sufficient for propositions, lemmas and theorems in the Appendix.
\begin{assumption}[$L^1$-modulus of the mechanism]\label{ass:modulus}
\begin{enumerate}[label=(\roman*),leftmargin=2.2em] \hfill
\item \emph{(envelope)} for some (hence every) $y_0\in S$ and every compact $K'\subseteq\Theta$,
\[
\int_{\Xs}\sup_{\theta\in K'}\dist\big(y_0,G(\theta,x)\big)\,\nu(dx)<\infty ;
\]
\item \emph{(continuity in $L^1(\nu)$)} for every $\theta\in\Theta$,
\[
\omega_\theta(\delta)\ :=\ \int_{\Xs}\ \sup_{\theta'\in\Theta:\ \|\theta'-\theta\|<\delta}\dist\big(G(\theta',x),G(\theta,x)\big)\ \nu(dx)\ \longrightarrow\ 0
\qquad(\delta\downarrow 0).
\]
\end{enumerate}
\end{assumption}
Assumption~\ref{ass:modulus}(i)
implies $\E\,\dist(y_0,Y_1)<\infty$, that is $P_\theta\in\PspOne(S)$, by taking $K'$ as a singleton. Assumption~\ref{ass:modulus}(ii) makes $\theta\mapsto P_\theta$ continuous in $\Wone$
(Lemma~\ref{lem:weak-cont}), which is equivalent to Assumption~2.2 of \citet{bernton2019parameter}.

\begin{assumption}[Identifiability]\label{ass:ident}
$\theta\ne\theta'\Rightarrow\Wone(P_\theta,P_{\theta'})>0$.
\end{assumption}

\begin{assumption}[Compact search set]\label{ass:search}
The estimation stage \eqref{eq:stage0} is solved over a compact set
$K\subseteq\Theta$ with $\theta_0\in K$.
\end{assumption}

Assumption~\ref{ass:search} implies Assumption~2.3 of
\citet{bernton2019parameter} on the search set, and it serves synthetic approximation accuracy. $\Pthm{\theta}$ approximates $P_\theta$
uniformly only on compacts (Lemma~\ref{lem:unif-synth}). In practice $K$ is the box over which
Algorithm~\ref{alg:estimation} searches, chosen to carry prior
mass close to one.

The recourse assumptions concern the merged atoms of Section~\ref{sec:mol_approx}
at the fixed mesh $\{\xb_j\}_{j\le N_1}$. Define the \emph{exceptional set}
\begin{equation}
\begin{aligned}
D\ :=\ &\Big\{\theta\in\Theta:\ \text{the fiber partition }\{I_\ell(\cdot)\}\text{ is not constant on any neighborhood of }\theta\Big\}\\
&\cup\ \bigcup_{j\le N_1}\Big\{\theta\in\Theta:\ G(\cdot,\xb_j)\text{ is discontinuous at }\theta\Big\}.
\end{aligned}
\label{eq:collision}
\end{equation}

\begin{assumption}[Local stability of the merged atoms]\label{ass:atoms}
The prior does not charge the exceptional set, $\Pi(D)=0$.
\end{assumption}

For $\theta_0\notin D$ there is a neighborhood $U$ of $\theta_0$ on which the fibers,
hence the number of atoms $L(\theta)=L(\theta_0)$ and their labels, are constant, and on
which each atom $z_\ell(\theta)=G(\theta,\xb_j)$, $j\in I_\ell(\theta_0)$, is continuous at
$\theta_0$. For $\theta\in U$ the cells $V_\ell(\theta)$ of \eqref{eq:wvor} are therefore defined with a common
number of atoms and common labels, and the following maps make sense.

\begin{assumption}[Continuity of the cell masses]\label{ass:boundary}
Almost surely under $\Pi$, for every $\ell\le L(\theta_0)$ the maps
\[
\theta\ \mapsto\ P_{\theta_0}\big(V_\ell(\theta)\big)
\qquad\text{and}\qquad
\theta\ \mapsto\ P_{\theta}\big(V_\ell(\theta)\big)
\]
are continuous at $\theta_0$.
\end{assumption}

Assumption~\ref{ass:boundary} is exactly what the recourse consistency theorem uses. The first map
controls the population target of the recourse weights, the second the model's own Voronoi masses.
It is not checked directly; instead, either of the following two conditions implies it
(Lemma~\ref{lem:suff}, Appendix~\ref{sec:proofs}). The first suits continuous observations and atoms that move with
$\theta$; the second suits lattice-valued observations, whose atoms do not move.

\begin{enumerate}[label=(5\alph*),leftmargin=3em]
\item\label{cond:5a} \emph{(Boundary-null cells.)} Almost surely under $\Pi$, $P_{\theta_0}$ does not charge the
bisectors of the merged atoms,
\[
P_{\theta_0}\Big(\textstyle\bigcup_{\ell\ne\ell'}\big\{y\in S:\ \dist\big(y,z_\ell(\theta_0)\big)=\dist\big(y,z_{\ell'}(\theta_0)\big)\big\}\Big)=0 .
\]
\item\label{cond:5b} \emph{(Fixed atoms on a discrete support.)} Almost surely under $\Pi$ there are a neighborhood
$U'$ of $\theta_0$ and a uniformly discrete set $\mathcal Y\subset S$, i.e.\
$\eta:=\inf\{\dist(y,y'):\ y\ne y'\in\mathcal Y\}>0$, such that
(i) $z_\ell(\theta)=z_\ell(\theta_0)$ for every $\ell$ and every $\theta\in U'$, and
(ii) $P_\theta(\mathcal Y)=1$ for every $\theta\in U'$.
\end{enumerate}

The verification of Assumptions~\ref{ass:modulus}--\ref{ass:boundary} for a location--scale family, for Poisson counts and for
Ornstein--Uhlenbeck paths is given in Appendix~\ref{sec:app-examples}. Table~\ref{tab:assumptions} summarizes which condition
gives Assumption~\ref{ass:boundary} in each example of the paper. Continuous observations whose atoms move with $\theta$ are
covered by \ref{cond:5a} and integer-valued observations are covered by \ref{cond:5b}.

\begin{table}[t]
\centering
\caption{Examples and the condition giving Assumption~\ref{ass:boundary}.}
\label{tab:assumptions}
\resizebox{\textwidth}{!}{%
\begin{tabular}{llll}
\hline
Example & Noise & Observation & Assumption~\ref{ass:boundary} via\\
\hline
Location--scale (Example~\ref{ex:loc_scale}, Section~\ref{sec:exp-gauss}) & $Z$ on $\R$ & $\R$ & \ref{cond:5a}\\
Poisson counts (Example~\ref{ex:poisson}) & Poisson process & $\N_0$ & \ref{cond:5b}\\
Segregating sites (Section~\ref{sec:exp-coal}) & Genealogy, Poisson clock & $\N_0$ & \ref{cond:5b}\\
Yule process (Section~\ref{sec:exp-yule}) & Poisson random measure & $\N_0^4$ & \ref{cond:5b}\\
Ornstein--Uhlenbeck (Example~\ref{ex:ou}, Section~\ref{sec:exp-ou}) & Brownian path & $L^2([0,T])$ & \ref{cond:5a}\\
\hline
\end{tabular}}
\end{table}

\section{Experiments}
\label{sec:experiments}

We illustrate the procedure on four models: a Gaussian location--scale family, the number of segregating sites under Kingman's coalescent, a pure birth process, and an Ornstein--Uhlenbeck diffusion. In
every example the noise is simulated in the coordinates in which the mechanism is written, and the recourse mesh is the $N_1$-medoid of a subsample of the synthetic
noise draws (Section~\ref{sec:mol_approx}). The estimation stage is Algorithm~\ref{alg:estimation}. For scalar observations the
objective is the exact $\Wone$ distance between the two empirical measures; for vector-valued observations
it is the sliced Wasserstein distance over the coordinate axes and $50$ random directions \citep{nadjahi2019asymptotic}. The quantization (pseudo)metric $\dist_Q$ of the mesh (Algorithm~\ref{alg:quantize}, with exponent $r=2$) and the metric $\dist$ of the recourse stage (Algorithm~\ref{alg:recourse}) are stated for each example below. Each held-out cell weight is compared with the model's own Voronoi mass
$w^{\mathrm{vor}}_\ell(\thn)$ through the exact pointwise binomial $95\%$ interval implied by
\eqref{eq:multinomial}; cells above the interval are reported as \emph{excess} and cells below it as
\emph{deficit}. In each example the structural parameter $\theta_0$ is a single draw from the prior,
and the prior test of the estimation stage is reported before the recourse. Table~\ref{tab:exp-summary} summarizes the four experiments.

\begin{table}[t]
\centering
\small
\caption{Summary of the experiments: noise law, quantization metric $\dist_Q$ of the mesh, recourse metric $\dist$, the
misspecification, and the $p$-value of the recourse Pearson test on one data set under correct specification and under the
misspecification.}
\label{tab:exp-summary}
\resizebox{\textwidth}{!}{%
\begin{tabular}{llllll}
\hline
Example & Noise & $\dist_Q$ & $\dist$ & Misspecification \\
\hline
Gaussian & $\mathcal N(0,1)$ & $|x-x'|$ & $|y-y'|$ & $t_2$; $5\%$ Gross errors \\
Segregating sites & Genealogy, clock & Euclidean, $(\log L_n,\log R)$ & $|y-y'|$ & Exponential growth \\
Yule & Poisson random measure & Euclidean, atom counts & Euclidean, $\R^4$ & Saturating birth rate \\
Ornstein--Uhlenbeck & Wiener measure & Discretized $L^2$ & Discretized $L^2$ & Concentrated noise & \\
\hline
\end{tabular}}
\end{table}

\subsection{Gaussian location--scale family}
\label{sec:exp-gauss}

\paragraph{Model and misspecifications.} The mechanism is
\[
Y=G(\theta,X)=\mu+\sigma X,\qquad X\sim\nu=\mathcal N(0,1),\qquad \theta=(\mu,\sigma^2),
\]
with the normal--inverse-gamma prior 
\[
\sigma^2\sim\mathrm{InvGamma}(3,2); \quad \mu\mid\sigma^2\sim\mathcal N(0,\sigma^2).
\]
The single draw is $\theta_0=(0.684,\,0.767)$. We use $n=5000$
observations split into $n_0=4000$ and $n_1=1000$, a synthetic sample of $m=10^4$ standard normal draws,
and a mesh of $N_1=100$ medoids of $M=2000$ of them, so that $n_1/N_1=10$. The quantization metric is
$\dist_Q(x,x')=|x-x'|$ on the noise space $\Xs=\R$, and the recourse metric is $\dist(y,y')=|y-y'|$ on $S=\R$. 

Three laws generate the data:
the model's own law $\mathcal N(\mu_\star,\sigma_\star^2)$ (\texttt{normal}); heavy tails,
$\mu_\star+\sigma_\star t_2$ (\texttt{t2}), and gross
errors, where $5\%$ of the observations are replaced by draws from
$\mathcal N(\mu_\star+10\sigma_\star,0.5^2)$ before the split (\texttt{contam}). 

\paragraph{Estimation stage and prior test.} Table~\ref{tab:gauss-est} compares the minimum-$\Wone$ estimate
with the maximum likelihood estimate. Under correct specification the two agree. Under $t_2$ the
sample standard deviation is dominated by extreme observations,
whereas the minimum-$\Wone$ estimate is less sensitive ($\tilde\sigma=1.27$ against $\hat\sigma=3.23$). Under contamination the maximum
likelihood estimate of $\sigma$ is off by $1.25$ and that of $\mu$ by $0.45$, about $16$ and $4.5$ times the
corresponding errors of the minimum-$\Wone$ fit. This demonstrates qualitative robustness of
Proposition~\ref{prop:robust}. 

\begin{table}[t]
\centering
\caption{Location--scale family: estimation stage. Minimum-$\Wone$ estimate (Algorithm~\ref{alg:estimation})
against maximum likelihood; truth $(\mu_\star,\sigma_\star)=(0.684,\,0.876)$.}
\label{tab:gauss-est}
\begin{tabular}{lcccc}
\hline
 & $\tilde\mu$ ($\Wone$) & $\hat\mu$ (MLE) & $\tilde\sigma$ ($\Wone$) & $\hat\sigma$ (MLE)\\
\hline
\texttt{normal} & 0.682 & 0.667 & 0.902 & 0.893\\
\texttt{t2}     & 0.717 & 0.784 & 1.269 & 3.234\\
\texttt{contam} & 0.784 & 1.134 & 0.952 & 2.123\\
\hline
\end{tabular}
\end{table}

\begin{figure}[t]
\centering
\includegraphics[width=\textwidth]{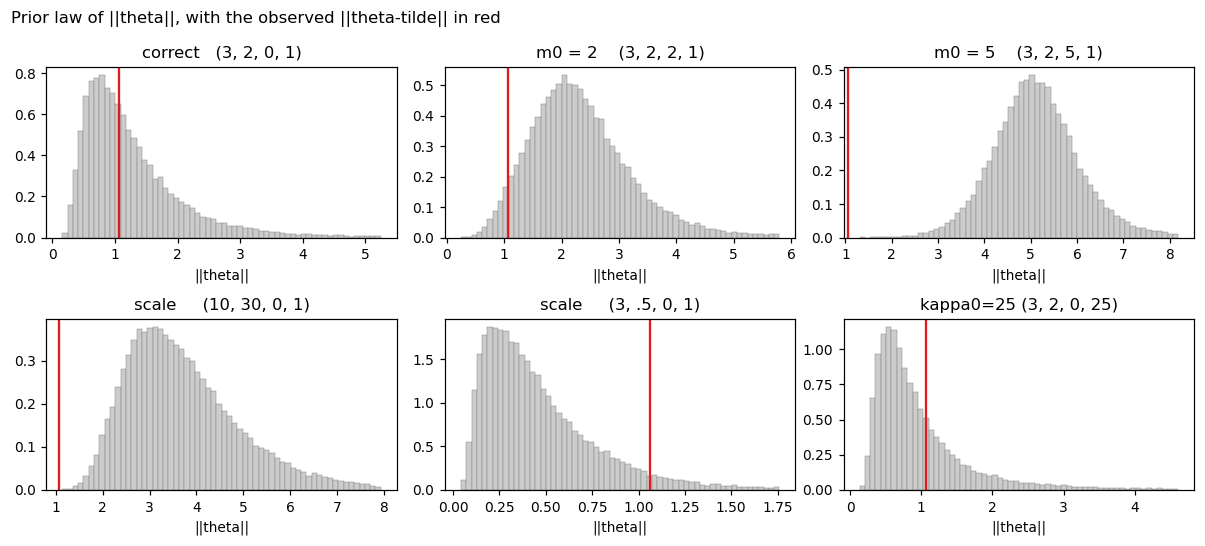}
\caption{Gaussian location--scale family: prior law of $\|\theta\|$, with the observed $\|\tilde\theta\|$ in red, under the
correct prior $(\alpha_0,\beta_0,m_0,\kappa_0)=(3,2,0,1)$ and five misspecified normal--inverse-gamma priors.}
\label{fig:gauss-prior}
\end{figure}

\paragraph{Prior misspecification.} The prior test of Section~\ref{sec:criticism} scores $\|\tilde\theta\|$ against
its law under a candidate prior. Under the prior that generated $\theta_0$, the fit to the \texttt{normal} data is
typical ($p=0.99$). We then score the same $\tilde\theta$ against five misspecified normal--inverse-gamma priors
$(\alpha_0,\beta_0,m_0,\kappa_0)$ (Figure~\ref{fig:gauss-prior}). Only the two gross misspecifications are rejected
at the $5\%$ level: a prior mean shifted to $m_0=5$ and a rescaled inverse-gamma with $(\alpha_0,\beta_0)=(10,30)$ (both $p<10^{-4}$). The milder misspecifications
are not detected: a shift to $m_0=2$ ($p=0.09$), a prior on $\sigma^2$ concentrated around $0.25$ ($p=0.13$), and a
prior with $\kappa_0=25$ ($p=0.61$). The prior test sees a single draw
$\theta_0$ through a single statistic, so it has power only when the prior is grossly misspecified, in the sense that
the realized $\tilde\theta$ falls in the far tail of the law of $\|\theta\|$. The last row shows that a change in the
prior spread of one coordinate can leave the law of $\|\theta\|$ almost unchanged.

\begin{figure}[t]
\centering
\includegraphics[width=\textwidth]{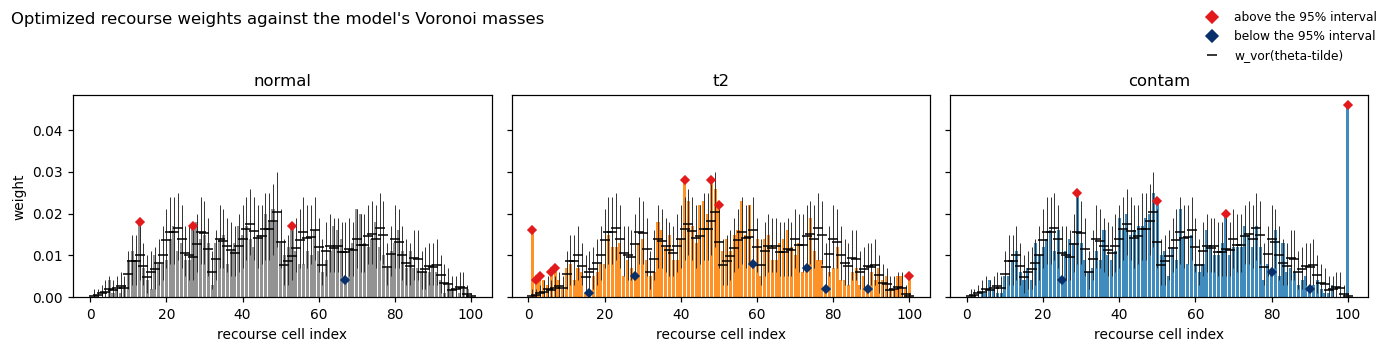}
\caption{Gaussian location--scale family: optimized recourse weights $\tilde w_\ell$ (bars) against the model's Voronoi
masses $w^{\mathrm{vor}}_\ell(\thn)$ (black ticks) with pointwise $95\%$ binomial intervals (whiskers), for the
\texttt{normal}, \texttt{t2} and \texttt{contam} data. Red diamonds: cells above the interval (excess); blue
diamonds: cells below it (deficit). Cells are ordered by their atom.}
\label{fig:gauss-weights}
\end{figure}

\begin{figure}[t]
\centering
\includegraphics[width=\textwidth]{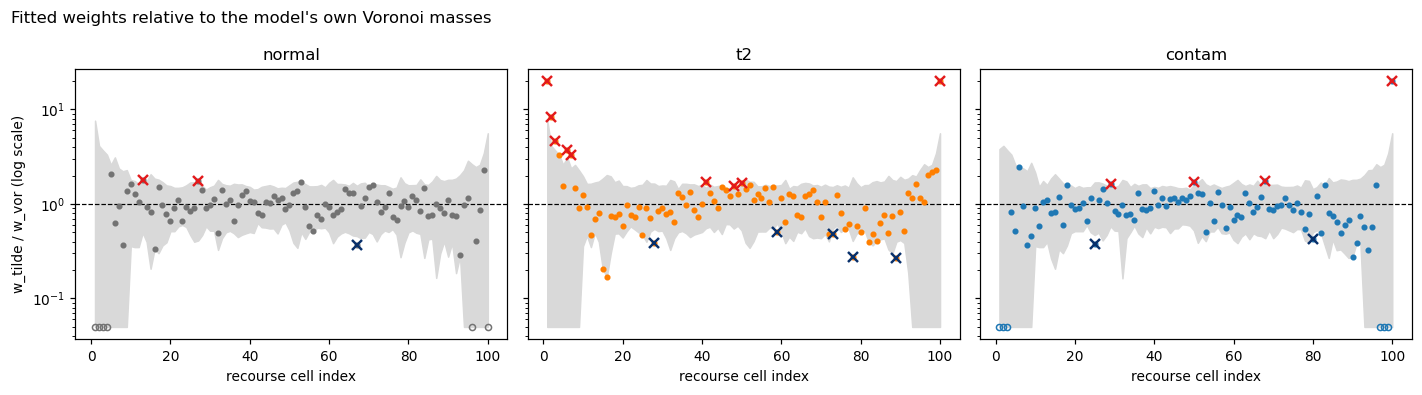}
\caption{Gaussian location--scale family: $\tilde w_\ell/w^{\mathrm{vor}}_\ell(\thn)$ on a logarithmic scale. Grey band:
pointwise $95\%$ null envelope at each case's $\thn$; red and blue crosses: cells above and below the band.}
\label{fig:gauss-ratio}
\end{figure}

\paragraph{Recourse stage.} Figure~\ref{fig:gauss-weights} shows the recourse weights against the model's
Voronoi masses and Figure~\ref{fig:gauss-ratio} their ratio on a logarithmic scale with a simulated $95\%$
null envelope. Under correct specification $4$ of the $100$ cells fall outside their pointwise interval,
against $3.05$ expected, and the Pearson test gives $p=0.59$. Under $t_2$ the two boundary cells carry about
$61$ and $28$ times their model mass. We view this as a signature of a noise law with heavier tails than the model. Under contamination the profile is one-sided. The top cell carries $257$ times its model mass. The Pearson test rejects decisively in both cases
($p=6\times10^{-26}$ and $6\times10^{-20}$).

\begin{figure}[t]
\centering
\includegraphics[width=0.5\textwidth]{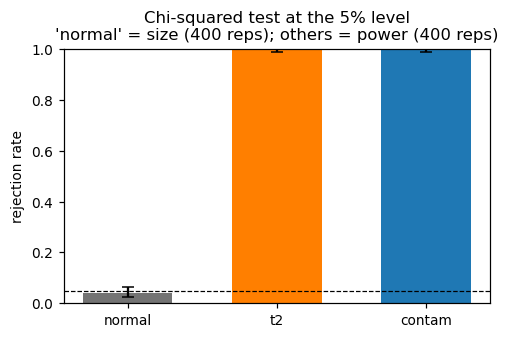}
\caption{Gaussian location--scale family: rejection rate of the recourse $\chi^2$ test at the $5\%$ level over $400$
replicates, with exact $95\%$ intervals; the \texttt{normal} bar is the size.}
\label{fig:gauss-power}
\end{figure}

\paragraph{Calibration and power.} Over $400$ replicate data sets of the whole pipeline, with the synthetic
sample and the mesh held fixed, the rejection rate at the $5\%$ level is $0.040$ (Monte Carlo standard error
$0.010$) under correct specification, and $1.00$ under both alternatives (Figure~\ref{fig:gauss-power}).

\clearpage
\subsection{Segregating sites under the coalescent}
\label{sec:exp-coal}

\paragraph{Model and misspecification.} A sample of $n=10$ sequences from a locus has a genealogy given by
Kingman's coalescent \citep{kingman1982coalescent}, and mutations fall on its branches as a Poisson process.
We write the noise as the coalescence exponentials $E_2,\dots,E_n$ and a unit-rate Poisson clock $N$ for the
mutations, and the mechanism as
\[
S_n=G(\theta,x)=N\big(\theta L_n/2\big),\qquad L_n=\sum_{k=2}^{n}\frac{2E_k}{k-1},
\]
the number of clock arrivals by internal time $\theta L_n/2$, so that $S_n\mid L_n\sim\mathrm{Poisson}(\theta L_n/2)$
\citep{watterson1975number}. Observations are counts at independent loci. The probability mass function of
$S_n$ is known \citep{tavare1984line}, so the Voronoi masses are exact. The mesh is the $N_1=300$-medoid of
$M=2500$ synthetic noise draws. The quantization metric $\dist_Q$ is the Euclidean distance between two standardized
summaries of a noise draw, the genealogy length $\log L_n$ and the clock rate $\log R$, $R=10/P_{10}$, where $P_{10}$ is the tenth arrival of the mutation clock. The quantization is a pseudometric in the space
\[
\left( (E_{2},\cdots,E_{10})^{\top}, (\epsilon_{1},\cdots)^{\top} \right) \in (0,\infty)^{9} \times (0,\infty)^{\mathbb{N}},
\]
where $\epsilon_{k} \stackrel{i.i.d.}{\sim} Exp(1)$ represents the waiting times in the unit-rate Poisson process. The recourse metric is
$\dist(y,y')=|y-y'|$ on the counts, $S=\R$. The prior is $\mathrm{LogNormal}(\log5,0.5^2)$ and $\theta_0=8.48$; $n_0=4000$, $n_1=1000$,
$m=10^4$. The misspecification is exponential population growth at rate $g=0.7$ \citep{slatkin1991pairwise}. Backwards in time $k$ lineages coalesce at rate $\binom k2e^{gt}$, which makes the genealogies more star-like
and their total length larger.

\paragraph{Estimation stage.} The minimum-$\Wone$ estimate agrees with Watterson's estimator in both cases
($8.36$ against $8.37$ under correct specification, $5.95$ against $5.94$ under growth), and the prior test finds
$\tilde\theta$ typical ($p=0.30$ and $0.73$; $p=0.29$ at $\theta_0$). 

\begin{figure}[t]
\centering
\includegraphics[width=0.9\textwidth]{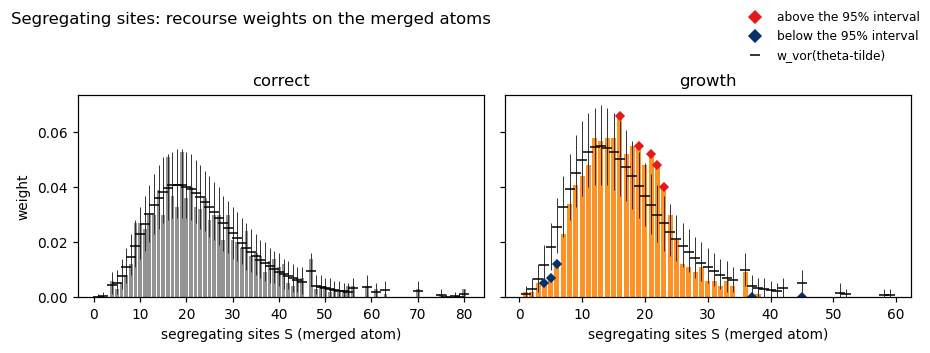}
\caption{Segregating sites: recourse weights on the merged count atoms against the exact Voronoi masses, for the
constant-size (\texttt{correct}) and exponential-growth (\texttt{growth}) data. Red and blue diamonds: excess and
deficit at the pointwise $95\%$ level.}
\label{fig:coal-weights}
\end{figure}

\begin{figure}[t]
\centering
\includegraphics[width=0.9\textwidth]{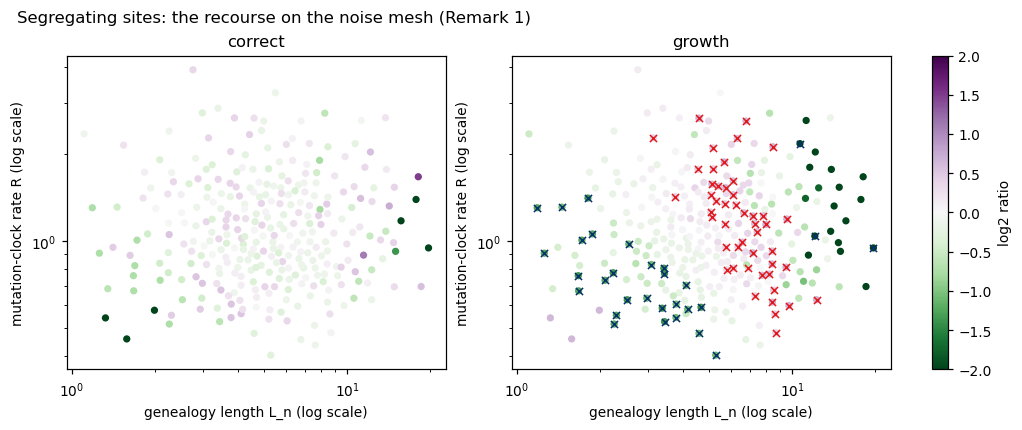}
\caption{Segregating sites: the recourse lifted to the noise mesh (Remark~\ref{rmk:fiber}). Each mesh point is
placed at its genealogy length $L_n$ and clock rate $R$ (log scales) and coloured by
$\log_2(\tilde w_\ell/w^{\mathrm{vor}}_\ell)$ of its fiber; red and blue crosses mark fibers with excess and deficit.
The fibers are bands of constant $L_nR$.}
\label{fig:coal-noise}
\end{figure}

\paragraph{Recourse stage.} The $300$ mesh points merge into $63$ count atoms under correct specification and
$46$ under growth (Figure~\ref{fig:coal-weights}). Under correct specification no atom is flagged and the
Pearson test gives $p=0.83$; under growth it gives $p=3\times10^{-9}$. The noise-space display of Remark~\ref{rmk:fiber} (Figure~\ref{fig:coal-noise}) shows how the misspecification
appears in the noise coordinates of the fitted model. The mesh points in the excess fibers have long
genealogies (median $L_n=6.5$), whereas those in the deficit fibers have short
genealogies ($L_n=3.0$). At the same time the display makes the
non-identifiability of Remark~\ref{rmk:fiber} explicit. The
data cannot tell a longer genealogy from a slower mutation clock and vice versa. The departure is reported along the bands, as a joint statement about tree length and mutation rate. 

\clearpage
\subsection{Yule process}
\label{sec:exp-yule}

\paragraph{Model and misspecification.} A population starts from $z_0=5$ individuals, each giving birth at rate
$\lambda$ and none dying. Each replicate is recorded at $t=1,2,3,4$, so $Y=(Z(1),\dots,Z(4))\in\R^4$ and
$\theta=\lambda$. By \eqref{eq:poisson-sde} with a single reaction channel,
\[
Z(t)=z_0+\int_0^t\!\!\int_0^\infty\mathbf 1\{u\le\lambda Z(s-)\}\,N(ds,du),
\]
and the noise is the Poisson sample path $N$, simulated on the window $[0,4]\times[0,150]$. An atom $(s,u)$ is a
candidate birth at time $s$, accepted when $u\le\lambda Z(s-)$. The mesh is the $150$-medoid of $1500$ of
$m=2000$ synthetic configurations. The quantization metric $\dist_Q$ is the Euclidean distance between the atom
counts of two configurations on a grid of rectangles of $[0,4]\times[0,32]$ ($8$ time bins of width $0.5$ times $5$
height bins with edges $0,2,4,8,16,32$), a pseudometric on the space of point configurations. The recourse metric is the
Euclidean distance on $S=\R^4$. Since the law of $(Z(1),\dots,Z(4))$ has no tractable form, the Voronoi masses
are frequencies over $40{,}000$ paths simulated from the fitted model. The prior is
$\lambda\sim\mathrm{LogNormal}(\log0.5,0.15^2)$, with $\lambda_0=0.576$; $n_0=4000$, $n_1=1000$. The data are generated
by Gillespie's direct method \citep{gillespie1977exact}. The misspecification is a state-dependent birth rate. In
\texttt{saturating} the per-capita birth rate is $\lambda_0K/(K+z)$ with $K=10$. Therefore, the growth is close to exponential while $z\ll K$ and close to linear in
time once $z\gg K$, whereas the Yule process has a constant per-capita rate and grows exponentially throughout.

\paragraph{Estimation stage.} Under correct specification $\tilde\lambda=0.571$, with prior $p$-value $0.37$ ($0.34$ at
$\lambda_0$). Under the saturating rate a single constant rate can only match the average growth over the four time
points, and the fit falls to $\tilde\lambda=0.297$, which the prior test flags as atypical ($p=0.0005$).

\begin{figure}[t]
\centering
\includegraphics[width=0.9\textwidth]{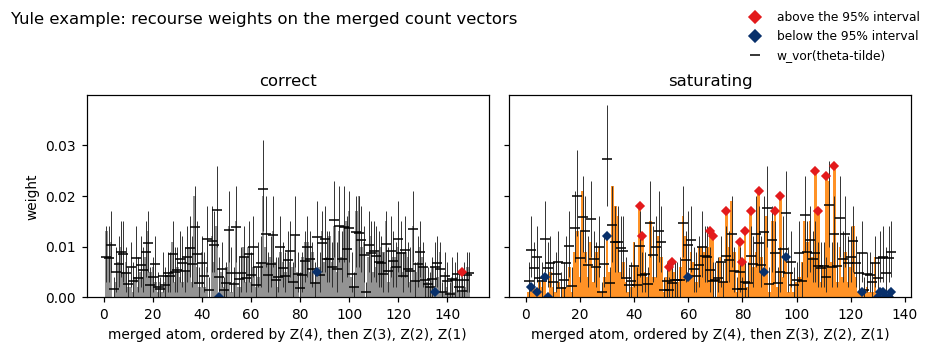}
\caption{Yule process: recourse weights on the merged count vectors, ordered by $Z(4)$ and then by the earlier counts,
for the \texttt{correct} and \texttt{saturating} data. Red and blue diamonds: excess and deficit at the pointwise
$95\%$ level.}
\label{fig:yule-weights}
\end{figure}

\begin{figure}[t]
\centering
\includegraphics[width=0.9\textwidth]{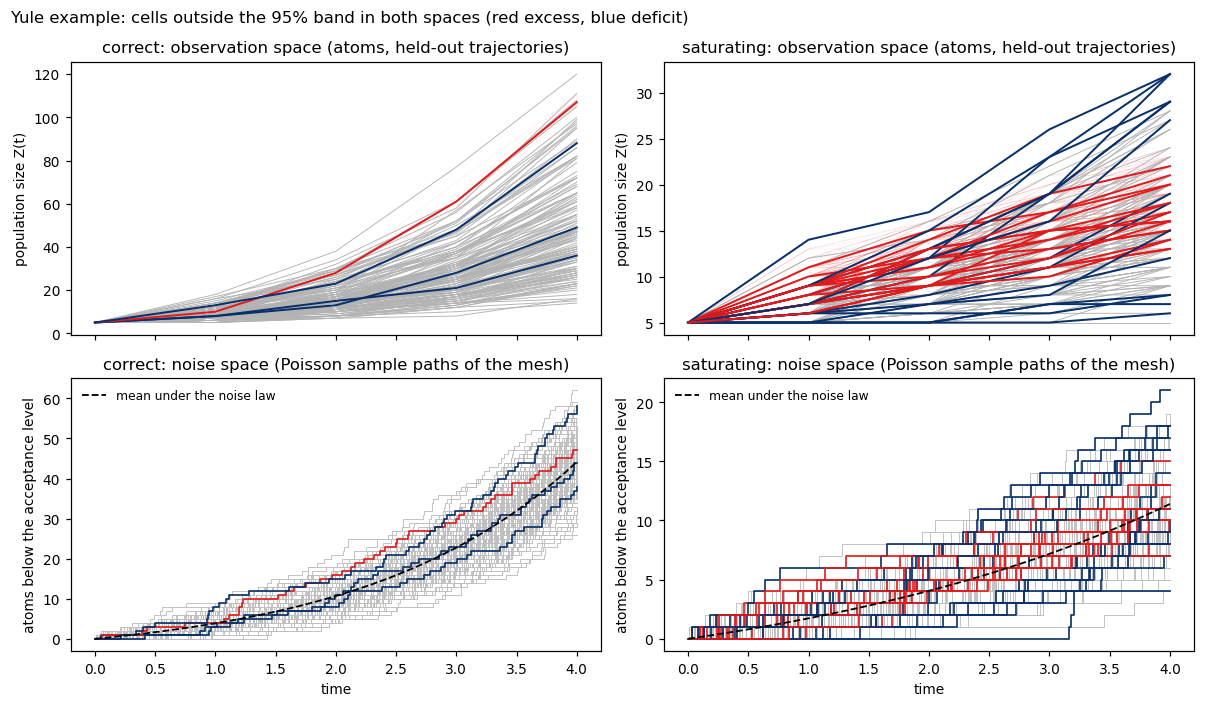}
\caption{Yule process: atoms drawn as count trajectories. Top: atoms in observation space. Bottom: the corresponding mesh paths in noise space. Red: excess; blue: deficit; grey: inside.}
\label{fig:yule-paths}
\end{figure}

\paragraph{Recourse stage.} The mesh merges into $149$ and $135$ distinct count vectors. Under correct specification
$4$ cells are flagged against $4.2$ expected and the Pearson test gives $p=0.34$; under the saturating rate $33$ cells
are flagged against $3.9$ expected and $p=2\times10^{-54}$ (Figure~\ref{fig:yule-weights}). The held-out trajectories grow too fast early and too slowly late for any Yule process.
Drawn as trajectories (Figure~\ref{fig:yule-paths}), the excess atoms are nearly linear paths and the deficit atoms
are the accelerating. Hence, the recourse locates the failure
in the shape of the growth curve, which is exactly what the state dependence of the birth rate changes. The mesh points are sample paths of
the Poisson random measure $N$, and each is drawn as the counting process
\[
N_{\tilde\lambda}(t)=\#\big\{(s,u)\in N:\ s\le t,\ u\le\tilde\lambda\,z_0e^{\tilde\lambda s}\big\}.
\]
In other words, $N$ is summarized by the number of atoms below the acceptance level of the fitted model along its mean path
$z_0e^{\tilde\lambda s}$. Note the observation y-axes are higher as $Z$ has a self-reinforcing growth rate $\lambda Z$ that the noise summary deliberately leaves out.

\clearpage
\subsection{Ornstein--Uhlenbeck diffusion}
\label{sec:exp-ou}

\paragraph{Model and misspecification.} The observation is the
Ornstein--Uhlenbeck path $dY=-aY\,dt+\sigma\,dW$, $Y(0)=0$ \citep{uhlenbeck1930theory}, written through its It\^o map
\[
Y=G(\theta,x)=\sigma\,(x-aK_ax),\qquad (K_ax)(t)=\int_0^te^{-a(t-u)}x(u)\,du,\qquad \theta=(a,\sigma),
\]
applied to a Brownian path $x$ on $[0,2]$ simulated on $200$ steps; $K_ax$ is evaluated exactly for the piecewise-linear
interpolation of the path, so no discretization scheme for the stochastic differential equation is involved. The mesh is the $80$-medoid of $2000$ of $m=10^4$ simulated
Brownian paths. The quantization metric $\dist_Q$ is the discretized $L^2$ distance, the Euclidean distance between the
values of two paths on the $200$-step grid, so the mesh points are Brownian paths and the atoms are OU paths. The
recourse metric is the Euclidean distance on $S=\R^{10}$, a discretized $L^2([0,2])$ distance between observed paths. The
prior is $a\sim\mathrm{LogNormal}(\log2,0.3^2)$, $\sigma\sim\mathrm{LogNormal}(0,0.3^2)$, with $\theta_0=(2.06,\,1.08)$;
$n_0=4000$, $n_1=1000$. The misspecification concerns the noise law. The data are $G(\theta_0,X)$ with $X$ the solution of
$dX=-1.5X\,dt+dW$, $X(0)=0$ (\texttt{concentrated\_noise}). The driving path moves locally like a Brownian motion but is
pulled back towards the origin, so it spreads less.

\paragraph{Estimation stage.} Under correct specification $\tilde\theta=(2.02,\,1.07)$ ($p=0.998$ in the prior test, $0.95$
at $\theta_0$). Under the concentrated noise the fitted model compensates by a much stronger mean reversion,
$\tilde\theta=(5.00,\,1.22)$, which the prior test flags as atypical ($p=0.002$).

\begin{figure}[t]
\centering
\includegraphics[width=0.9\textwidth]{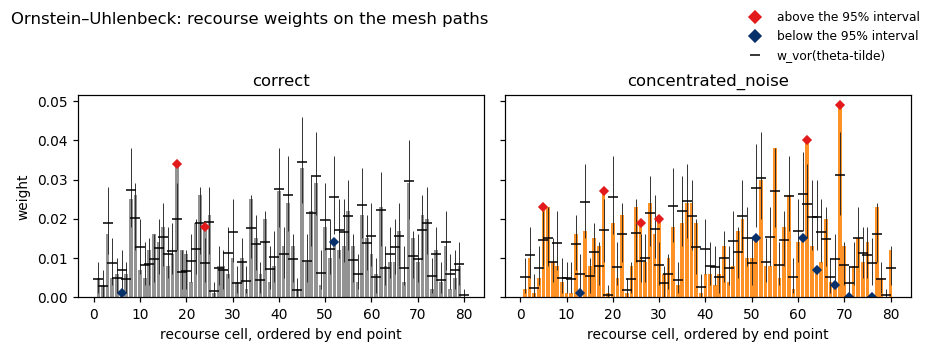}
\caption{Ornstein--Uhlenbeck diffusion: recourse weights on the mesh paths, ordered by the end point $Z(T)$ of the atom,
for the \texttt{correct} and \texttt{concentrated\_noise} data. Red and blue diamonds: excess and deficit.}
\label{fig:ou-weights}
\end{figure}

\begin{figure}[t]
\centering
\includegraphics[width=\textwidth]{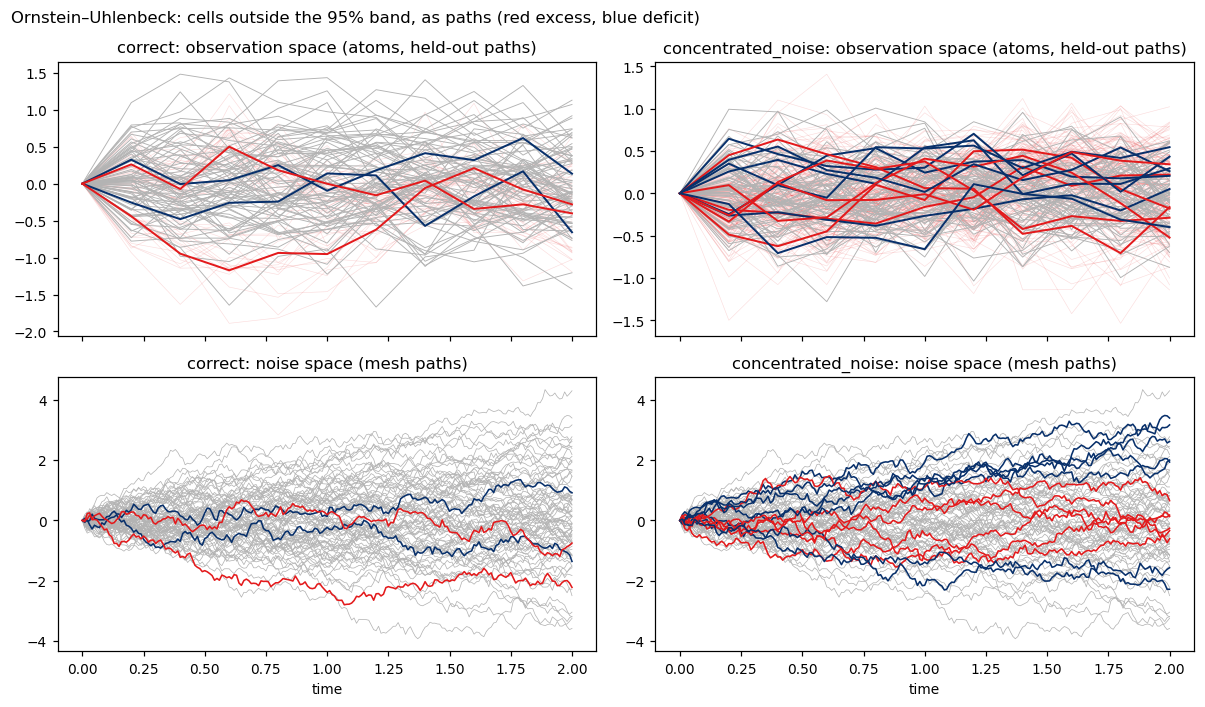}
\caption{Ornstein--Uhlenbeck diffusion: cells outside the pointwise $95\%$ band, as paths. Top: atoms in observation space. Bottom: the corresponding mesh paths in noise space. Red: excess; blue: deficit; grey: inside.}
\label{fig:ou-paths}
\end{figure}

\paragraph{Recourse stage.} Under correct specification $4$ of the $80$ cells are flagged against $2.7$ expected and the
Pearson test gives $p=0.04$; this is a single data set, on which a $p$-value below $0.05$ under the null has probability
$0.05$, and the Gaussian example shows that the test holds its level. Under the concentrated noise $13$ cells are flagged
against $2.6$ expected and $p=2\times10^{-12}$ (Figure~\ref{fig:ou-weights}). Because the noise is a Brownian path, the
diagnostic is read off paths (Section~\ref{sec:closedform}). In noise space (Figure~\ref{fig:ou-paths}, bottom row) the
deficit cells are exactly the mesh paths with large excursions and the excess cells the paths that stay near the origin. The recourse recovers that the driving noise is more concentrated than the standard Brownian motion.

\clearpage

\section{Computation}
\label{sec:computation}

The estimation stage minimizes $\Wone(\Pxo,\Pthm{\theta})$ over
$\theta\in K$ and the recourse stage minimizes $\Wone(\Pxi,P_{\thn,\bw})$ over
$\bw\in\Delta^{L-1}$. The first compares two empirical measures, the data
and the synthetic sample; the second compares an empirical measure with a molecular measure on the merged atoms.

\subsection{Evaluating the estimation-stage objective}
\label{sec:W1eval}

Write $a=(a_t)$ for the weights of $\Pxo$ and $b=(b_i)$ for
those of $\Pthm{\theta}$ (at the recourse stage, source $\Pxi$ and
$b=\bw$), and let $C_{ti}=\dist(y_t,G(\theta,\tilde x_i))$ be the
ground-cost matrix. Then
\begin{equation}
\Wone\big(\Pxo,\Pthm{\theta}\big)=\min_{\pi\ge0}\ \ip{C}{\pi}
\ \ \text{s.t.}\ \ \pi\mathbf 1=a,\ \pi^\top\mathbf 1=b,
\label{eq:kanto}
\end{equation}
which is solved by the transportation (network) simplex or any linear-programming
solver \citep{peyre2019computational}. A standard implementation is \citet{flamary2021pot}.

\paragraph{High-dimensional approximation: sliced $\Wone$.} For $s>1$ the program \eqref{eq:kanto} has $n_0m$ variables and must be
solved at every point visited by Algorithm~\ref{alg:estimation}. We therefore use, at the estimation stage only, the sliced
distance
\[
\mathrm{SW}_1\big(\Pxo,\Pthm{\theta}\big)\ :=\ \frac1P\sum_{p=1}^{P}\Wone\big((\pi_p)_{\#}\Pxo,\ (\pi_p)_{\#}\Pthm{\theta}\big),
\qquad \pi_p(y)=\ip{v_p}{y},
\]
over a fixed set of unit directions $v_1,\dots,v_P$ drawn once together with the synthetic noise (in our experiments the
coordinate axes and $50$ random directions). Each term is a one-dimensional distance between two sorted samples,
$\Wone(\mu,\nu)=\int_\R|F_\mu-F_\nu|$.

\subsection{Estimation stage: a derivative-free search over $\theta$}
\label{sec:solver0}

The objective $F_{n_0}(\theta):=\Wone(\Pxo,\Pthm{\theta})$ is in general not differentiable in
$\theta$. We therefore use
a single derivative-free search, Algorithm~\ref{alg:estimation}. It requires only evaluations of $F_{n_0}$, each of which is one of the
transport problems of Section~\ref{sec:W1eval}. The grid phase locates the global basin, which matters because $F_{n_0}$ may be multimodal; the
Nelder--Mead phase \citep{nelder1965simplex} refines within it. The search is embarrassingly parallel. 

\subsection{Recourse mesh: computing the $N_1$-medoid}
\label{sec:quantize}

The mesh of Section~\ref{sec:mol_approx} is the $N_1$-medoid \eqref{eq:medoid}, which minimizes the empirical distortion
\begin{align*}
    e_r(\Gamma;\nu_M)^r=M^{-1}\sum_{i\le M}\min_{j\le N_1}\dist_Q(x^{(i)},\xb_j)^r,\qquad \nu_M:=M^{-1}\textstyle\sum_{i\le M}\delta_{x^{(i)}},
\end{align*}
over meshes of $N_1$ sample points. Algorithm~\ref{alg:quantize} combines a
randomized seeding with a guarantee and a local improvement that never increases the distortion.

\subsection{Recourse-stage algorithm}
\label{sec:solver1}

With $\thn$ frozen, the recourse requires no optimization. Equation~\eqref{eq:voronoi} \citep[Lem.~4.9]{PflugPichler2014} says the optimal
weights are nearest-atom counts and the value is the mean nearest-atom
distance, so the entire stage is a merge followed by a nearest-neighbour pass over the
held-out block $\xi_{1}$. The output of
Algorithm~\ref{alg:recourse} is exactly the recourse optimum.

\begin{algorithm}[t]
\caption{Estimation stage: derivative-free search for Equation~\eqref{eq:stage0}.}
\label{alg:estimation}
\begin{algorithmic}[1]
\Require estimation block $\xi_0$; synthetic noise $\tilde x_{1:m}$ drawn once from $\nu$;
search box $K$; grid $\theta^{(1)},\dots,\theta^{(G)}\in K$ of mesh width $h$; number of
starts $r\le G$; tolerance $\epsilon>0$
\For{$g=1,\dots,G$ \textbf{in parallel}}
\State $F_g\gets\Wone\big(\Pxo,\Pthm{\theta^{(g)}}\big)$ \quad (Section~\ref{sec:W1eval})
\EndFor
\State let $\vartheta^{(1)},\dots,\vartheta^{(r)}$ be the $r$ grid points with the smallest values $F_g$
\For{$c=1,\dots,r$ \textbf{in parallel}}
\State run Nelder--Mead on $F_{n_0}$ from $\vartheta^{(c)}$, with $F_{n_0}:=+\infty$ outside $K$,
until the simplex values differ by less than $\epsilon$; record its best vertex $\hat\vartheta^{(c)}$
\EndFor
\State \Return $\thn$, the point with the smallest objective among
$\{\theta^{(g)}\}_{g\le G}\cup\{\hat\vartheta^{(c)}\}_{c\le r}$
\end{algorithmic}
\end{algorithm}

\begin{algorithm}[t]
\caption{Recourse mesh: approximate $N_1$-medoid of simulated noise.}
\label{alg:quantize}
\begin{algorithmic}[1]
\Require synthetic noise $\tilde x_{1:m}$ of the estimation stage (or a sampler for $\nu$); mesh size $N_1$; subsample
size $M\le m$; quantization (pseudo)metric $\dist_Q$; exponent $r\ge1$; tolerance $\epsilon>0$; seed
\State draw a random subsample $x^{(1)},\dots,x^{(M)}$ of $\tilde x_{1:m}$ \quad (reduce $M$ if the search below is too slow)
\State \emph{(seeding, $k$-medoids\texttt{++})}\quad $\xb_1\gets x^{(i)}$ with $i$ uniform on $\{1,\dots,M\}$; for $j=2,\dots,N_1$,
$\xb_j\gets x^{(i)}$ with probability proportional to $\min_{j'<j}\dist_Q(x^{(i)},\xb_{j'})^r$
\Repeat
\State \emph{(assignment)}\quad assign each $x^{(i)}$ to its nearest mesh point, ties broken by lowest index
\State \emph{(medoid update)}\quad replace each $\xb_j$ by the sample point of its cell that minimizes
$\sum_{i\in C_j}\dist_Q(x^{(i)},\cdot)^r$
\State \emph{(swap)}\quad replace a mesh point $\xb_j$ by a sample point $x^{(i)}$ whenever this lowers the empirical distortion
\Until{the empirical distortion decreases by less than $\epsilon$}
\State $q_j\gets$ frequency of cell $C_j$ among the synthetic draws not in the subsample (or among fresh draws from $\nu$)
\State \Return mesh $\{\xb_j\}_{j\le N_1}$, masses $\{q_j\}_{j\le N_1}$, and the seed
\end{algorithmic}
\end{algorithm}

\begin{algorithm}[t]
\caption{Recourse stage: closed-form solution of Equation~\eqref{eq:stage1-primal} and the calibrated test.}
\label{alg:recourse}
\begin{algorithmic}[1]
\Require held-out block $\xi_1$; fitted $\thn$; mesh $\{\xb_j\}_{j=1}^{N_1}$ from Algorithm~\ref{alg:quantize}
\State raw atoms $G(\thn,\xb_j)$ for $j=1,\dots,N_1$
\State \emph{(merge)}\quad sort or hash the raw atoms and group equal values into the fibers $I_\ell$ of \eqref{eq:fibers}, labelled by smallest index; obtain distinct atoms $z_1,\dots,z_L$
\State build a nearest-neighbour index on $\{z_\ell\}$ 
\For{$t\in\xi_1$}
\State $\ell(t)\gets\argmin_{\ell}\dist(y_t,z_\ell)$, ties broken by lowest
label; \quad $d_t\gets\dist\big(y_t,z_{\ell(t)}\big)$
\EndFor
\State $\tilde w_\ell\gets\#\{t:\ \ell(t)=\ell\}/n_1$ for $\ell=1,\dots,L$; \quad
$Q(\thn)\gets n_1^{-1}\sum_{t\in\xi_1}d_t$
\State $w^{\mathrm{vor}}_\ell(\thn)\gets P_{\thn}(V_\ell(\thn))$ for
$\ell=1,\dots,L$, by simulation from the fitted model (for atoms of $P_{\thn}$, the empirical frequency of $\{G(\thn,X)=z_\ell\}$, cf.\ \eqref{eq:wvor-discrete})
\State pool cells with small expected count $n_1 w^{\mathrm{vor}}_\ell(\thn)$; compute $T_{n_1}$ of \eqref{eq:chisq} and its $\chi^2$ $p$-value
\State \Return weights $\tilde\bw$, value $Q(\thn)$, test result, and the diagnostic
profile $\tilde w_\ell$ against $w^{\mathrm{vor}}_\ell(\thn)$ on the merged atoms
\end{algorithmic}
\end{algorithm}

\clearpage

\section{Discussion}
\label{sec:beyond}

We have cast model criticism for structural models $Y=G(\theta,X)$ as a sequential two-stage stochastic program
on a sample split. The estimation stage fits the structure with the noise law held at its assumed value, by a robust
minimum-Wasserstein estimator computed by sample average approximation. The recourse stage freezes the structure and lets
the held-out data reweight the atoms of the fitted model. The recourse has a closed form, so the procedure requires no
choice of test statistic. The same object, the vector of held-out cell frequencies on the atoms of the fitted model,
serves both as a calibrated test and as a diagnostic that says where the model fails. The four experiments show the
diagnostic separating symmetric tail defects from one-sided gross errors, exposing population growth through the shape of
the law of the number of segregating sites, locating a state-dependent birth rate in the shape of the growth curve, and
recovering that a driving noise is more concentrated than Brownian motion.

\citet{ratmann2009model} proposed model criticism within approximate
Bayesian computation The ABC tolerance is replaced by an error term on each summary statistic, and the posterior of these
errors, computed jointly with the parameters, signals which summaries the model fails to reproduce. That approach shares
our setting, models that can be simulated but whose likelihood is not available, and our aim of reporting \emph{where} a
model fails. It differs in two respects. First, its diagnostic is expressed in the summary statistics chosen by the analyst,
whereas ours is expressed on the atoms of the fitted model, i.e.\ on regions of the noise space, and requires no summary. The
price is that the diagnostic is resolved only up to the fibers of the fitted mechanism (Remark~\ref{rmk:fiber}). Second,
the posterior of the errors is a Bayesian summary without calibration, whereas the split makes the recourse
counts conditionally multinomial and the Pearson test calibrated.

We conclude with limitations. The theory requires i.i.d.\ replicates, so a single long trajectory of a process is not covered. The recourse identifies the noise law only on the $\sigma$-algebra generated by the
fitted mechanism, so a departure is reported jointly for all noise values in a fiber as demonstrated in the coalescent example in Section \ref{sec:exp-coal}. The resolution of the
diagnostic is bounded by the mesh, and refining the mesh trades resolution for noisier cell weights through the ratio
$n_1/N_1$. The reference masses are plug-ins at $\tilde\theta$, which adds to Pearson's statistic an amount of order $n_1/n_0$. Therefore, the estimation block should be several times larger than the held-out block. Finally, since the structure, the noise law and
the prior are all entangled, a misspecified structure can be absorbed into the noise law and reported as a noise-law departure.

\section{Code Availability}
The code is publicly available on GitHub at \\ \url{https://github.com/RenYiW/Model-Criticism-via-Stochastic-Optimization.git}.

\section{Acknowledgments}
The author thanks Professors David Blei and Simon Tavaré for their helpful comments on this manuscript. RYW acknowledges support from the Herbert and Florence Irving Institute for Cancer Dynamics at Columbia University.

\section{Competing Interests}
The author declares no competing interests.

\clearpage
\appendix
\section{Proofs}
\label{sec:proofs}

\subsection{Estimation stage}
\label{sec:proofs-estimation}

\begin{proposition}[Wasserstein strong law]\label{prop:W1-SLLN}
Let Assumption~\ref{ass:modulus}(i) hold and let
$Y_1,Y_2,\dots\stackrel{iid}{\sim}P_{\theta_0}$. Then
\begin{align*}
\Prob\Big\{\ \lim_{m\to\infty}\Wone\big(\tfrac1m\textstyle\sum_{t=1}^m\delta_{Y_t},\,P_{\theta_0}\big)=0\ \Big\}=1.
\end{align*}
Consequently, $\Wone(\Pxo,P_{\theta_0})\to 0$ as
$n_0\to\infty$ and $\Wone(\Pxi,P_{\theta_0})\to0$ as
$n_1\to\infty$, both almost surely. The same holds with $P_{\theta_0}$ replaced by
$P_\theta$ for any fixed $\theta\in\Theta$ and the sample replaced by $G(\theta,\tilde x_{1:m})$.
\end{proposition}

\begin{proof}[Proof of Proposition~\ref{prop:W1-SLLN}]
This is \citet[Thm.~6.9]{Villani2009} combined with the strong law of large numbers
for the empirical measure and its first moment. Quantitative rates are available under higher moments \citep{FournierGuillin2015}.
\end{proof}

\begin{lemma}[$\Wone$-continuity of the model map]\label{lem:weak-cont}
Under Assumption~\ref{ass:modulus}, for all $\theta,\theta'\in\Theta$ with $\|\theta'-\theta\|<\delta$,
\[
\Wone(P_\theta,P_{\theta'})\ \le\ \int_{\Xs}\dist\big(G(\theta,x),G(\theta',x)\big)\,\nu(dx)\ \le\ \omega_\theta(\delta).
\]
In particular $\theta_k\to\theta$ implies $\Wone(P_{\theta_k},P_\theta)\to0$, hence $P_{\theta_k}\Rightarrow P_\theta$.
\end{lemma}

\begin{proof}[Proof of Lemma~\ref{lem:weak-cont}]
The law of $(G(\theta,X),G(\theta',X))$, $X\sim\nu$, is a coupling of $P_\theta$ and $P_{\theta'}$. Since
\begin{align*}
    \Wone(P_\theta,P_{\theta'}) = \inf_{\pi \in \Pi(\nu, \nu)} \int_{\mathcal{X}\times \mathcal{X}} \dist\big(G(\theta,x_{1}),G(\theta',x_{2})\big) \,\pi(d x_1,d x_2),
\end{align*}
the first inequality holds. The second is the definition of
$\omega_\theta$. Continuity follows from Assumption~\ref{ass:modulus}(ii), and $\Wone$ convergence implies
weak convergence \citep[Thm.~6.9]{Villani2009}.
\end{proof}

\begin{lemma}[Continuity of the exact objective and the synthetic error]\label{lem:cont-obj}
For every fixed $Q\in\PspOne(S)$, the map $\theta\mapsto\Wone(Q,P_{\theta})$ is continuous on
$\Theta$ under Assumption~\ref{ass:modulus}, and for every $\theta$ and $m$,
\[
\big|\Wone(Q,\Pthm{\theta})-\Wone(Q,P_\theta)\big|\ \le\ \Wone\big(P_\theta,\Pthm{\theta}\big).
\]
\end{lemma}

\begin{proof}[Proof of Lemma~\ref{lem:cont-obj}]
Both statements follow from the reverse triangle inequality,
\[
|\Wone(Q,\mu)-\Wone(Q,\mu')|\le\Wone(\mu,\mu'),
\]
with $(\mu,\mu')=(P_\theta,P_{\theta'})$ and
Lemma~\ref{lem:weak-cont} for the first, and $(\mu,\mu')=(\Pthm{\theta},P_\theta)$ for the second.
\end{proof}

\begin{lemma}[Uniform synthetic approximation]\label{lem:unif-synth}
Let Assumption~\ref{ass:modulus} hold and let $K\subseteq\Theta$ be compact. Then
\[
\Delta_m\ :=\ \sup_{\theta\in K}\Wone\big(P_\theta,\Pthm{\theta}\big)\ \longrightarrow\ 0
\qquad\text{almost surely as } m\to\infty .
\]
\end{lemma}

\begin{proof}[Proof of Lemma~\ref{lem:unif-synth}]
For $\theta\in K$ and $\delta>0$, define
\begin{align*}
    \Delta_{\theta,\delta}(x):=\sup_{\theta'\in\Theta,\,\|\theta'-\theta\|<\delta}\dist\big(G(\theta',x),G(\theta,x)\big),
\end{align*}
such that $\int\Delta_{\theta,\delta}\,d\nu=\omega_\theta(\delta)$ is finite for small $\delta$, this is possible due to Assumption~\ref{ass:modulus}(ii). For $\theta'$ satisfying
$\|\theta'-\theta\|<\delta$, the triangle inequality, Lemma~\ref{lem:weak-cont}, and the coupling of
$\Pthm{\theta}$ and $\Pthm{\theta'}$ that pairs $G(\theta,\tilde x_i)$ with $G(\theta',\tilde x_i)$ give
\begin{align*}
\Wone\big(P_{\theta'},\Pthm{\theta'}\big)
&\le\Wone(P_{\theta'},P_\theta)+\Wone\big(P_\theta,\Pthm{\theta}\big)+\Wone\big(\Pthm{\theta},\Pthm{\theta'}\big)\\
&\le\omega_\theta(\delta)+\Wone\big(P_\theta,\Pthm{\theta}\big)+\frac1m\sum_{i=1}^m\Delta_{\theta,\delta}(\tilde x_i).
\end{align*}
Note the same $\tilde x_i$ generate both synthetic samples.
By Proposition~\ref{prop:W1-SLLN} applied to $P_\theta$ and the strong law of large numbers applied to
$\Delta_{\theta,\delta}(\tilde x_i)$, almost surely
\[
\limsup_{m\to\infty}\ \sup_{\|\theta'-\theta\|<\delta}\Wone\big(P_{\theta'},\Pthm{\theta'}\big)\ \le\ 2\,\omega_\theta(\delta).
\]
Fix $\epsilon>0$. By Assumption~\ref{ass:modulus}(ii), each $\theta\in K$ has a $\delta_\theta>0$ with
$\omega_\theta(\delta_\theta)\le\epsilon/2$. By compactness finitely many balls
$\{\theta'\mid\|\theta'-\theta_q\|<\delta_{\theta_q}\}$ cover $K$. Intersecting the finitely many almost sure events gives
$\limsup_m\Delta_m\le\epsilon$ almost surely. Letting $\epsilon\downarrow0$ along a sequence proves the claim.
This is a bracketing Glivenko--Cantelli argument in the sense of \citet[Thm.~2.4.1]{vandervaart1996weak}.
\end{proof}

\begin{lemma}[Inverse continuity at $\theta_0$]\label{lem:inv-cont}
Under Assumptions~\ref{ass:modulus}, \ref{ass:ident} and \ref{ass:search}, if $\theta_k\in K$ and
\begin{align*}
    \Wone(P_{\theta_0},P_{\theta_k})\to 0, \quad\text{then}\quad \|\theta_k-\theta_0\|\to 0.
\end{align*}
\end{lemma}

\begin{proof}[Proof of Lemma~\ref{lem:inv-cont}]
Since $K$ is compact, any subsequence of $(\theta_k)$ has a further subsequence
$\theta_{k_i}\to\theta^\ast\in K$. By
Lemma~\ref{lem:weak-cont}, $\Wone(P_{\theta_{k_i}},P_{\theta^\ast})\to0$, so
$\Wone(P_{\theta_0},P_{\theta^\ast})\le\Wone(P_{\theta_0},P_{\theta_{k_i}})+\Wone(P_{\theta_{k_i}},P_{\theta^\ast})\to0$,
which implies $\theta^\ast=\theta_0$ by Assumption~\ref{ass:ident}. The subsequence argument then gives $\theta_k\to\theta_0$.
\end{proof}

\begin{theorem}[Strong consistency of $\thn$]\label{thm:consistency}
Suppose Assumptions~\ref{ass:modulus}--\ref{ass:search} hold, let the synthetic sample size satisfy
$m=m(n_0)\to\infty$, and let $\thn$ be any measurable $\varepsilon_{n_0}$-approximate minimizer of
\eqref{eq:stage0} with $\varepsilon_{n_0}\downarrow0$. Then
\[
\big\|\thn-\theta_0\big\|\ \longrightarrow\ 0
\qquad\text{almost surely as } n_0\to\infty .
\]
\end{theorem}

\begin{proof}[Proof of Theorem~\ref{thm:consistency}]
Work on the intersection of the almost sure events of Proposition~\ref{prop:W1-SLLN} and
Lemma~\ref{lem:unif-synth}, and write $\Delta:=\Delta_{m(n_0)}$. By Lemma~\ref{lem:cont-obj}, for every
$\theta\in K$, $|\Wone(\Pxo,\Pthm{\theta})-\Wone(\Pxo,P_\theta)|\le\Delta$. Since $\theta_0, \thn\in K$,
\begin{align*}
\Wone\big(P_{\theta_0},P_{\thn}\big)
&\le\Wone(P_{\theta_0},\Pxo)+\Wone\big(\Pxo,\Pthm{\thn}\big)+\Delta
\le\Wone(P_{\theta_0},\Pxo)+\Wone\big(\Pxo,\Pthm{\theta_0}\big)+\varepsilon_{n_0}+\Delta\\
&\le2\,\Wone(P_{\theta_0},\Pxo)+\varepsilon_{n_0}+2\Delta\ \longrightarrow\ 0 .
\end{align*}
Lemma~\ref{lem:inv-cont} concludes $\big\|\thn-\theta_0\big\|\ \longrightarrow\ 0$. Equivalently, $\thn$ is an
$(\varepsilon_{n_0}+2\Delta)$-approximate minimizer of $\theta\mapsto\Wone(\Pxo,P_\theta)$ over $K$, so that
the conclusion also follows from \citet[Thm.~2.1]{bernton2019parameter} in its $\varepsilon_n$-argmin form.
\end{proof}

\begin{proposition}[Qualitative robustness at the model]\label{prop:robust}
Let Assumptions~\ref{ass:modulus}--\ref{ass:search} hold, let $Q_k\in\PspOne(S)$ with
$\Wone(Q_k,P_{\theta_0})\to0$, and let $\theta_k\in K$ satisfy
$\Wone(Q_k,P_{\theta_k})\le\inf_{\theta\in K}\Wone(Q_k,P_\theta)+\varepsilon_k$ with $\varepsilon_k\to0$. Then
$\theta_k\to\theta_0$. The conclusion persists if $P_\theta$ is replaced by $\Pthm{\theta}$ in the
definition of $\theta_k$, with $m=m(k)\to\infty$.
\end{proposition}

\begin{proof}[Proof of Proposition~\ref{prop:robust}]
Since $\theta_0\in K$,
\[
\Wone(P_{\theta_0},P_{\theta_k})\le\Wone(P_{\theta_0},Q_k)+\Wone(Q_k,P_{\theta_k})
\le\Wone(P_{\theta_0},Q_k)+\Wone(Q_k,P_{\theta_0})+\varepsilon_k\to0,
\]
and Lemma~\ref{lem:inv-cont} gives $\theta_k\to\theta_0$. For the synthetic version, Lemma~\ref{lem:cont-obj}
shows that an $\varepsilon_k$-minimizer of $\theta\mapsto\Wone(Q_k,\Pthm{\theta})$ over $K$ is an
$(\varepsilon_k+2\Delta_{m(k)})$-minimizer of $\theta\mapsto\Wone(Q_k,P_\theta)$, with $\Delta_m$ of
Lemma~\ref{lem:unif-synth}, and the same argument applies, exactly as in the proof of
Theorem~\ref{thm:consistency}.
\end{proof}

\subsection{Recourse stage}
\label{sec:proofs-recourse}

\begin{lemma}[Conditional concentration of the recourse weights]\label{lem:GC}
Let the data be i.i.d.\ from $P_{\theta_0}$ and let the regime \eqref{eq:regime} hold. Write
$p_\ell:=P_{\theta_0}\big(V_\ell(\thn)\big)$, $\ell\le L(\thn)$, for the population masses of the recourse cells. Then,
almost surely,
\[
\max_{\ell\le L(\thn)}\ \big|\tilde w_\ell-p_\ell\big|\ \longrightarrow\ 0
\qquad(n\to\infty).
\]
\end{lemma}

\begin{proof}[Proof of Lemma~\ref{lem:GC}]
Given $\Fzero$, the parameter $\thn$, the merged atoms and hence the cells 
\[
V_1(\thn),\dots,V_{L(\thn)}(\thn)
\]
are fixed
Borel sets partitioning $S$, and $\xi_1$ is an i.i.d.\ sample from $P_{\theta_0}$ independent of $\Fzero$. Hence
$n_1\tilde w_\ell\mid\Fzero\sim\mathrm{Binomial}(n_1,p_\ell)$ as in Eq.~\eqref{eq:multinomial}. Hoeffding's inequality
\citep{hoeffding1963probability} with a union bound over $\ell\le L(\thn)\le N_1$ gives, for every $\epsilon>0$,
\[
\Prob\Big(\max_{\ell\le L(\thn)}|\tilde w_\ell-p_\ell|>\epsilon\ \Big|\ \Fzero\Big)\ \\
\le \sum_{\ell\le L(\thn)}\Prob\Big(|n_1\tilde w_\ell- n_1 p_\ell|>n_1\epsilon\ \Big|\ \Fzero\Big) \\
\le 2N_1e^{-2n_1\epsilon^2}.
\]
Under regime \eqref{eq:regime},
\[
n_0 \wedge n_1 \to \infty; \quad \frac{n_1}{\log(n)} \to \infty.
\]
Hence, $n_1(n) \ge(1/\epsilon^2)\log n$ for all large $n$, and
\[
\sum_n2N_1e^{-2n_1(n)\epsilon^2}\le C+\sum_n2N_1n^{-2}<\infty.
\]
The
Borel--Cantelli lemma gives $\limsup_n\max_\ell|\tilde w_\ell-p_\ell|\le\epsilon$ almost surely. Intersecting over
$\epsilon=1/q$, $q\in\N$, proves the claim. 
\end{proof}

\begin{lemma}[Sufficient conditions for continuity of the cell masses]\label{lem:suff}
Let Assumptions~\ref{ass:modulus} and \ref{ass:atoms} hold and let $\theta_0\notin D$. If either \ref{cond:5a} or
\ref{cond:5b} holds at $\theta_0$, then for every $\ell\le L(\theta_0)$ the maps
$\theta\mapsto P_{\theta_0}(V_\ell(\theta))$ and $\theta\mapsto P_\theta(V_\ell(\theta))$ are continuous at $\theta_0$.
\end{lemma}

\begin{proof}[Proof of Lemma~\ref{lem:suff}]
Since $\theta_0 \notin D$, let $U$ be a neighborhood of $\theta_0$ on which the fiber partition is constant, we have
\[
I_{\ell}(\theta) = I_{\ell}(\theta_0), \quad \forall \theta \in U, \ell \leq L(\theta_{0}).
\]
Let $\theta_k\to\theta_0$ with $\theta_k\in U$. By
the continuity part of Assumption~\ref{ass:atoms}, 
\[
z_\ell(\theta_k)=G(\theta_k,\xb_j)\to z_\ell(\theta_0), \quad \forall j\in I_\ell(\theta_0), \ell \leq L(\theta_0).
\]

\emph{Under \ref{cond:5a}.} Let $y$ lie off every bisector. That is,
\[
y \notin \{\dist(y,z_\ell(\theta_0))=\dist(y,z_{\ell'}(\theta_0)), \quad \forall \ell'\ne\ell.
\]
The inequalities
$\dist(y,z_\ell(\theta_0))\lessgtr\dist(y,z_{\ell'}(\theta_0))$ are all strict. Since the atoms converge, these strict
inequalities hold for $z_{1:L}(\theta_k)$ for all large $k$, and
\[
\mathbf 1_{V_\ell(\theta_k)}(y)\to\mathbf 1_{V_\ell(\theta_0)}(y).
\]
By \ref{cond:5a} this holds
$P_{\theta_0}$-almost everywhere, and dominated convergence gives 
\[
P_{\theta_0}(V_\ell(\theta_k))\to P_{\theta_0}(V_\ell(\theta_0)).
\]
For the second map write $P_{\theta_k}(V_\ell(\theta_k))=\nu\{x:\ G(\theta_k,x)\in V_\ell(\theta_k)\}$. By
Lemma~\ref{lem:weak-cont},
\[
\int\dist\big(G(\theta_k,x),G(\theta_0,x)\big)\,\nu(dx)\to0,
\]
so every subsequence has a further
subsequence along which $G(\theta_k,x)\to G(\theta_0,x)$ for $\nu$-almost every $x$. By \ref{cond:5a},
\[
\nu \left(\{x \mid G(\theta_0,x) \text{ off the bisectors of } z_{1:L}(\theta_0)\} \right) = 1.
\]
The strict inequalities above
are preserved for the pair $(G(\theta_k,x),z_{1:L}(\theta_k))$, so
\[
\mathbf 1\{G(\theta_k,x)\in V_\ell(\theta_k)\}\to\mathbf 1\{G(\theta_0,x)\in V_\ell(\theta_0)\}.
\]
Dominated convergence
along the sub-subsequence, and the subsequence principle, give
\[
P_{\theta_k}(V_\ell(\theta_k))\to P_{\theta_0}(V_\ell(\theta_0)).
\]

\emph{Under \ref{cond:5b}.} For large $k$, $\theta_k\in U\cap U'$, where by (i) the atoms, and hence the cells
\[
V_\ell(\theta_k)=V_\ell(\theta_0)=:V_\ell,
\]
are fixed sets. The first map is therefore eventually constant. For the
second, write 
\[
p_\theta(y):=P_\theta(\{y\}), \quad \forall y \in \mathcal Y.
\]
The function
$f_y(z):=\max\{0,\,1-2\dist(z,y)/\eta\}$ is bounded and Lipschitz, equals $1$ at $y$ and vanishes at every other point of
$\mathcal Y$, so by (ii) $p_\theta(y)=\int f_y\,dP_\theta$ for $\theta\in U'$. By Lemma~\ref{lem:weak-cont},
$P_{\theta_k}\Rightarrow P_{\theta_0}$, hence $p_{\theta_k}(y)\to p_{\theta_0}(y)$ for every $y\in\mathcal Y$. Since
$\sum_{y\in\mathcal Y}p_\theta(y)=1$ for $\theta\in U'$, Scheff\'e's lemma \citep{scheffe1947useful} applied on $\mathcal Y$
with counting measure gives $\sum_{y\in\mathcal Y}|p_{\theta_k}(y)-p_{\theta_0}(y)|\to0$, and therefore
\[
\big|P_{\theta_k}(V_\ell)-P_{\theta_0}(V_\ell)\big|\ \le\ \sum_{y\in\mathcal Y}\big|p_{\theta_k}(y)-p_{\theta_0}(y)\big|\ \longrightarrow\ 0 .
\]
\end{proof}

\begin{theorem}[Recourse weight consistency]\label{thm:stage1-weights}
Let Assumptions~\ref{ass:modulus}--\ref{ass:boundary} hold for the fixed mesh $\{\xb_j\}_{j\le N_1}$, and let $\thn$ be as in
Theorem~\ref{thm:consistency}. Then, in the regime \eqref{eq:regime},
$L(\thn)=L(\theta_0)=:L$ almost surely for all $n_0$ large enough, and for each $\ell\le L$,
\[
\tilde w_\ell\ \longrightarrow\ w^{\mathrm{vor}}_\ell(\theta_0),
\qquad
w^{\mathrm{vor}}_\ell(\thn)\ \longrightarrow\ w^{\mathrm{vor}}_\ell(\theta_0),
\]
and $\Wone\big(\sum_\ell\tilde w_\ell\delta_{z_\ell(\thn)},\ \sum_\ell w^{\mathrm{vor}}_\ell(\theta_0)\delta_{z_\ell(\theta_0)}\big)\to0$.
\end{theorem}

\begin{proof}[Proof of Theorem~\ref{thm:stage1-weights}]
By Assumption~\ref{ass:atoms}, $\theta_0\notin D$ almost surely; let $U$ be a neighborhood of $\theta_0$ on
which the fiber partition is constant. By Theorem~\ref{thm:consistency}, almost surely $\thn\in U$ for all large
$n_0$, so $L(\thn)=L$ and the labels of $\thn$ and $\theta_0$ agree. In addition, $z_\ell(\thn)=G(\thn,\xb_j)$ for any
$j\in I_\ell(\theta_0)$, so $z_\ell(\thn)\to z_\ell(\theta_0)$ by the continuity part of Assumption~\ref{ass:atoms}.

Fix $\ell$ and split
\begin{align*}
    |\tilde w_\ell-w^{\mathrm{vor}}_\ell(\theta_0)|\leq \big|\tilde w_\ell-P_{\theta_0}(V_\ell(\thn))\big| + |P_{\theta_0}(V_\ell(\thn))-P_{\theta_0}(V_\ell(\theta_0))|.
\end{align*}
The first term tends to zero almost surely by Lemma~\ref{lem:GC}. The second tends to zero because
$\thn\to\theta_0$ and $\theta\mapsto P_{\theta_0}(V_\ell(\theta))$ is continuous at $\theta_0$ by
Assumption~\ref{ass:boundary}. Likewise 
\[
w^{\mathrm{vor}}_\ell(\thn)=P_{\thn}(V_\ell(\thn))\to P_{\theta_0}(V_\ell(\theta_0))
=w^{\mathrm{vor}}_\ell(\theta_0)
\]
by the continuity of the second map in Assumption~\ref{ass:boundary}.

By the triangle inequality, coupling atom $\ell$ with atom $\ell$ in the first term and
bounding $\Wone$ by the diameter times the total variation in the second,
\[
\Wone\Big(\sum_\ell\tilde w_\ell\delta_{z_\ell(\thn)},\sum_\ell w^{\mathrm{vor}}_\ell\delta_{z_\ell(\theta_0)}\Big)
\le\max_\ell\dist\big(z_\ell(\thn),z_\ell(\theta_0)\big)+\mathrm{diam}\{z_{1:L}(\theta_0)\}\sum_\ell|\tilde w_\ell-w^{\mathrm{vor}}_\ell(\theta_0)|\to0 .
\]
\end{proof}

\section{Verification of the assumptions for the examples}
\label{sec:app-examples}

This appendix verifies Assumptions~\ref{ass:modulus}--\ref{ass:boundary} for three families; Table~\ref{tab:assumptions}
summarizes the results.

\begin{example}[Location--scale family]\label{ex:loc_scale}
Let $S=\R$, $\Theta=\R\times(0,\infty)$, $\theta=(\mu,\sigma)$, and
\[
Y\ =\ \mu+\sigma Z,
\]
where $Z\in L^1$ is a continuous random variable with mean zero. That is, $\Xs=\R$, $\nu$ is the law of $Z$ and
$G(\theta,x)=\mu+\sigma x$. For Assumption~\ref{ass:modulus}, if $\|\theta'-\theta\|<\delta$ then
\[
|G(\theta',x)-G(\theta,x)|\le\delta\,(1+|x|),
\]
so $\omega_\theta(\delta)\le\delta\,(1+\E|Z|)$, and on a compact $K'$ the
envelope is bounded by $c_{K'}(1+|x|)$, which is $\nu$-integrable. Assumption~\ref{ass:ident} holds since $\mu=\E Y$ and $\sigma=\E|Y-\mu|/\E|Z|$. $G(\theta,\cdot)$ is strictly increasing and
continuous in $\theta$, so distinct mesh points have distinct images, the fibers are singletons for every $\theta$,
$D=\emptyset$, and Assumption~\ref{ass:atoms} holds. Finally, $P_{\theta_0}$ is absolutely continuous, so \ref{cond:5a} and
hence Assumption~\ref{ass:boundary} hold.
\end{example}

\begin{example}[Poisson counts]\label{ex:poisson}
Let $S=\R$, $\Theta=(0,\infty)$, and let the noise be a unit-rate Poisson process on $[0,\infty)$, recorded by its
arrival times $x=(P_1,P_2,\dots)$, $0<P_1<P_2<\cdots$; the noise space is a space of increasing sequences, which is
Polish, and $\nu$ is the law of the arrival times. The mechanism counts the arrivals up to time $\lambda$,
\[
G(\lambda,x)\ =\ N_x(\lambda)\ :=\ \#\{k:\ P_k\le\lambda\},
\]
so that $P_\lambda=\mathrm{Poisson}(\lambda)$; equivalently $G(\lambda,\cdot)$ is the generalized inverse
$F_\lambda^{-1}(U)$ of a uniform $U$ in law. For Assumption~\ref{ass:modulus}, $\lambda\mapsto N_x(\lambda)$
is nondecreasing, so for $\delta<\lambda$
\[
\omega_\lambda(\delta)\ \le\ \E\big[N(\lambda+\delta)-N(\lambda-\delta)\big]\ =\ 2\delta,
\]
and on $K'\subseteq[a,b]$ the envelope is $N_x(b)$, with $\E N(b)=b<\infty$. Assumption~\ref{ass:ident} holds since
$\Wone(P_\lambda,P_{\lambda'})\ge|\lambda-\lambda'|$ (test function $y\mapsto y$). 

For a fixed mesh point $\xb_j$, the map
$\lambda\mapsto G(\lambda,\xb_j)$ is piecewise constant and jumps exactly at the arrival times of $\xb_j$, so the fiber
partition changes only on the countable set
\[
D\ \subseteq\ \bigcup_{j\le N_1}\{P_1(\xb_j),P_2(\xb_j),\dots\}.
\]
This set is Lebesgue-null, hence $\Pi$-null whenever $\Pi$ is absolutely continuous, and Assumption~\ref{ass:atoms}
holds. Off $D$ every merged atom is locally constant, and $P_\lambda(\N_0)=1$ for every $\lambda$, so
condition~\ref{cond:5b} holds with $\mathcal Y=\N_0$ and $\eta=1$. 
\end{example}

\begin{example}[Ornstein--Uhlenbeck process]\label{ex:ou}
Let $\Xs=C_0([0,T])$ with the supremum norm, let $\nu$ be Wiener measure, let $S=L^2([0,T])$, and let
$\theta=(a,\sigma)\in\Theta=(0,\infty)^2$. Define, for every $x\in\Xs$,
\[
G(\theta,x)\ =\ \sigma\,(x-aK_ax),\qquad (K_ax)(t):=\int_0^te^{-a(t-u)}x(u)\,du .
\]
Integration by parts gives $G(\theta,W)(t)=\sigma\int_0^te^{-a(t-u)}\,dW(u)$, so $P_\theta$ is the law of the
Ornstein--Uhlenbeck path started at $0$, a centred Gaussian measure with covariance kernel
\[
\tfrac{\sigma^2}{2a}\big(e^{-a|t-u|}-e^{-a(t+u)}\big).
\]
Distinct $\theta$ give distinct kernels, so
Assumption~\ref{ass:ident} holds. On $L^2([0,T])$ the operator $K_a$ has norm at most $T$ and
$\|K_a-K_{a'}\|\le T^2|a-a'|$, so on a compact $K'\subset\Theta$ there is $c_{K'}$ with
\[
\|G(\theta,x)-G(\theta',x)\|_{L^2}\le c_{K'}\|\theta-\theta'\|\,\|x\|_{L^2} \text{ and } \|G(\theta,x)\|_{L^2}\le c_{K'}\|x\|_{L^2}.
\]
Since $\|x\|_{L^2}\le\sqrt T\|x\|_\infty$ and $\E\|W\|_\infty<\infty$, Assumption~\ref{ass:modulus} holds with
$\omega_\theta(\delta)\le c\,\delta$. 
The Volterra operator $I-aK_a$ is injective on $L^2([0,T])$ and distinct elements of $C_0([0,T])$ are distinct in
$L^2$, so distinct mesh points have distinct images, the fibers are singletons, $G(\cdot,\xb_j)$ is
continuous, $D=\emptyset$, and Assumption~\ref{ass:atoms} holds. Finally the covariance operator of $P_{\theta_0}$ is
$\sigma_0^2B C_W B^*$ with $B=I-a_0K_{a_0}$ and $C_W$ the injective Brownian covariance operator, which is
injective; hence \ref{cond:5a} holds. 
\end{example}

\bibliography{mybib}

\bibliographystyle{abbrvnat}

\end{document}